\documentclass[12pt]{article}
\usepackage{hyperref}

\usepackage[vcentermath]{youngtab}
\usepackage{color}

\def\be{\begin{equation}}
\def\ee{\end{equation}}
\def\ba{\begin{eqnarray}}
\def\ea{\end{eqnarray}}

\def\de{\right}
\def\si{\left}
\def\ap{\a {\rm '}}

\def\p{\partial}
\def\nb{\nonumber}

\def\beq{\begin{equation}}
\def\eeq{\end{equation}}
\def\bea{\begin{eqnarray}}
\def\eea{\end{eqnarray}}

\def\e{\epsilon}
\def\h{\eta}

\def\z{\zeta}
\def\m{\mu}
\def\n{\nu}
\def\r{\rho}
\def\s{\sigma}
\def\th{\theta}
\def\t{\tau}
\def\f{\varphi}
\def\a{\alpha}
\def\b{\beta}
\def\d{\delta}
\def\l{\lambda}

\def\w{\omega}

\def\la{\langle}
\def\ra{\rangle}
\usepackage[centertags]{amsmath}
\usepackage{amsfonts} 
\usepackage{amssymb}
\usepackage{amsthm}
\numberwithin{equation}{section} 

\begin{document}

\begin{titlepage}
\hfill \hbox{NORDITA-2015-123}
\vskip 0.1cm
\hfill \hbox{QMUL-PH-15-18}
\vskip 1.5cm
\begin{center}
{\Large \bf A microscopic description of absorption in high-energy string-brane collisions}
 
  \vskip 1.0cm {\large Giuseppe
D'Appollonio$^{a}$, Paolo Di Vecchia$^{b, c}$,
Rodolfo Russo$^{d}$, \\
Gabriele Veneziano$^{e, f}$ } \\[0.7cm]
{\it $^a$ Dipartimento di Fisica, Universit\`a di Cagliari and 
INFN\\ Cittadella
Universitaria, 09042 Monserrato, Italy}\\
{\it $^b$ The Niels Bohr Institute, University of Copenhagen, Blegdamsvej 17, \\
DK-2100 Copenhagen {\O}, Denmark}\\
{\it $^c$ Nordita, KTH Royal Institute of Technology and Stockholm University, \\Roslagstullsbacken 23, SE-10691 Stockholm, Sweden}\\
{\it $^d$ Queen Mary University of London, Mile End Road, E1 4NS London, United Kingdom}\\
{\it $^e$ Coll\`ege de France, 11 place M. Berthelot, 75005 Paris, France}\\
{\it $^f$Theory Division, CERN, CH-1211 Geneva 23, Switzerland}
\end{center}
\begin{abstract}
We study the collision of a highly energetic light closed string off a stack of D$p$-branes at (sub)string-scale impact parameters and in a regime justifying a perturbative treatment. 
Unlike at larger impact parameters --where elastic scattering and/or tidal excitations dominate-- here absorption of the closed string by the brane system,
with the associated excitation of open strings living on it, becomes important. As a first step, we study this phenomenon at the disk level, in which the energetic closed string turns into a single heavy open string at rest whose particularly simple  properties are described.
\end{abstract}
\end{titlepage}

\tableofcontents

\section{Introduction}
\label{intro}

The dynamics of strings in the background of a collection of  D$p$-branes
provides an excellent framework to 
address the problem of string dynamics in curved spacetimes. It also 
underlies many important developments in our understanding of quantum gravity, most notably the gauge-gravity
duality \cite{Maldacena:1997re, Witten:1998qj}.

In a series of relatively recent papers~\cite{D'Appollonio:2010ae,  D'Appollonio:2013hja, D'Appollonio:2013rsa, D'Appollonio:2015gpa} 
we have addressed the problem of the high-energy collision of a closed string off a configuration of   parallel D$p$-branes. By suitably playing with the various parameters characterizing the process, the number $N$ of branes, the energy $E$ and impact parameter $b$ of the collision, as well as the string coupling $g_s$, we could identify~\cite{D'Appollonio:2010ae} a region in parameter space inside which closed string loops can be safely neglected and, consequently, there is no closed string production or gravitational bremsstrahlung, a considerable simplification. Even within this region there are several
interesting regimes to study.

There is a {\it weak-gravity regime} 
(corresponding to  very large impact parameters compared to the other length scales in the problem) in which string-size effects can be neglected and general-relativity expectations are recovered  in terms  of gravitational deflection and Shapiro time delay. In this regime the resulting S-matrix satisfies elastic unitarity.

There is also a {\it string-size-corrected weak-gravity regime} 
 (corresponding to somewhat smaller impact parameters) in which tidal excitations of the incoming closed string become important or even 
dominant~\cite{D'Appollonio:2010ae}. In this regime we are still able to provide an exactly unitary S-matrix, but unitarity now works in an enlarged Hilbert space containing excited closed strings besides the incoming one. 
In~\cite{D'Appollonio:2013hja, D'Appollonio:2013rsa} (see also~\cite{Black:2011ep, Bianchi:2011se}) the microscopic structure of this S-matrix was analyzed in much detail.

There is finally a  {\it strong-gravity regime} 
(corresponding to a gravitational radius of the effective $p$-brane 
geometry $R_p$ larger than $b$) in which the closed string is captured by the brane system. Physically this is  the most interesting situation since, in the QFT limit, one expects information about the initial state to be lost in the capture (cf. the fall into a potential well in quantum mechanics).  By contrast, string theory should again be able to give a unitary (i.e. information preserving) S-matrix by providing a microscopic description of the (open string) excitations induced on the branes by the  absorbed closed string.

The study of this process is in general very hard. It can be  simplified, however, under the 
assumption
that $R_p$, while comparable or even larger than $b$, is still smaller than the string length parameter 
$l_s$, enhanced  by a square root of the logarithm of the energy, as we will discuss in more detail
in the rest of the paper~\footnote{In a previous paper~\cite{ D'Appollonio:2015gpa} we have studied, precisely in this 
regime, the elastic scattering of a closed string 
 for what concerns the resolution of the causality issue recently raised by Camanho et al.~\cite{Camanho:2014apa}. 
In this paper we look at a complementary aspect, the absorption of the closed string resulting in the above-mentioned production of open string excitations of the brane system.}.  When this is the case
the eikonal resummation of the leading terms (in energy) of the higher-order string amplitudes~\cite{Amati:1987wq, Amati:1987uf, D'Appollonio:2010ae}
should give a correct representation of the dynamics for every value of
the impact parameter, all the way down to $b=0$. Although in this limit a geometric interpretation of the brane background is lacking,
the dynamics of the string-brane system remains extremely rich and interesting. 

The absorption
process is expected to lead to a state consisting of a very complicated (yet quantum mechanically pure)  linear superposition of multi-open-string excitations of the brane system. Therefore, we may regard the problem at hand as being very close, in spirit, to the famous information puzzle arising from the formation and evaporation of a black hole from a pure initial state. 

The study of  a similar process was attempted before in the context of string-string collisions~\cite{Amati:1987wq, Amati:1987uf} where
 the regime analogous to the one considered here, corresponds to taking the gravitational radius $R_S \equiv 2 G_N \sqrt{s}$ to be smaller than $l_s$ but possibly larger than $b$. Although black hole formation is not supposed to happen in such a regime, the final state was argued~\cite{Amati:1987wq, Amati:1987uf, Veneziano:2004er} to have many features in common with the one expected from an evaporating black hole.
Only a rough description of the final state (basically just keeping track of the number of final strings) was obtained in~\cite{Amati:1987wq, Amati:1987uf, Veneziano:2004er}. The hope is that, in the case of string-brane collisions, one should be able to go much further in the microscopic description of the final state.

Our final aim is to arrive at a unitary S-matrix describing  both the tidal excitation and the absorption of the energetic incoming closed string. 
In this paper we shall take a first but  important step in this direction by studying in detail the process at  tree (i.e. disk) level. In this approximation  the energetic closed string is absorbed through the formation of a {\it single} massive open string attached to the brane system. In this paper we will  study the detailed properties of this highly excited state at the quantum level. In a forthcoming one \cite{classicalclop} we will be able to give a simple and intuitive explanation of such properties by considering a closed-string brane collision in a kinematical regime allowing for the formation of an open string at the classical level. Furthermore, when the conditions for the classical closed to open transition are not met it will be possible to perform a semiclassical analysis which confirms qualitatively the results described in this paper.

The main result of this paper is an explicit  microscopic description of the massive open
string created by the absorption of an arbitrary closed string. The open string belongs to the 
$n$-th level of the string spectrum, with $n \sim \ap E^2$ fixed by the energy of the closed string,
and as the energy increases we are exploring higher and higher levels of the spectrum. 
 Since the covariant methods are not very
suitable to deal with generic excited states, we will work in the light-cone gauge 
and derive the form of the open state by taking 
the high-energy limit of the light-cone closed-open 
vertex~\cite{Cremmer:1973ig, Clavelli:1973uk, Green:1983hw, Shapiro:1987gq, Green:1994ix}.

The dimensionality of the Hilbert space of possible final states
is exponentially large and one would expect that 
the massive open string would have a very complex representation in a generic basis. 
Instead, we find that by choosing a natural basis for the process, a light-cone gauge aligned with the direction
of large momentum, detailed calculations become possible leading to an extremely simple representation of the final state. 

As already mentioned, the detailed understanding of the absorption process at tree level is only the first step
in the construction of a unitary S-matrix. 
In order to achieve this, one needs to take into account higher (open string) loops.
We shall present elsewhere~\cite{clopunitary} how  to generate an explicitly unitary S-matrix in a suitable narrow-resonance approximation.
 
This paper is organized as follows. In Section 2 we review the string-brane elastic scattering amplitudes in the
Regge limit, presenting a general formula for arbitrary
external closed states using the eikonal operator and the Reggeon vertex operator. 
We then discuss the $s$-channel factorization of the amplitude and its imaginary part, which is 
the relevant quantity for describing the absorption of the closed string through the excitation of open strings living on the brane system. 

In Section 3 we introduce the closed-open transition amplitudes and the closed-open light-cone vertex. The latter is essential
to work with  arbitrary external closed and open states. The form of the closed-open vertex in the high-energy limit is derived in 
Section 4. These results are used in Section 5 to derive a simple
and explicit form for the open state created on the brane system by the absorption of a closed string. We analyze in turn
the absorption of a tachyon, of a  massless state and of a state belonging to the first massive level  and finally give
a formula for a generic closed string. As a test of our results we show how to reconstruct the imaginary part of the
elastic amplitude in impact parameter space. 
In Section 6 we present our conclusions.

We collected some additional material in two Appendices. In the first we compare 
the closed-open transition amplitudes given by the light-cone vertex and by the covariant methods
for states 
belonging to the lowest levels. In the second
we provide some technical details on the evaluation of the imaginary part of the disk using the closed-open vertex.

\section{Elastic scattering of closed strings by D-branes}
\label{scattsec}

When a closed string propagates in the presence of a stack of $N$ coincident D$p$-branes,
the simplest new processes that can occur at leading order in perturbation theory 
are its scattering (elastic 
or inelastic) off the D$p$-branes and its absorption by the D$p$-branes. Both processes are given by a 
two-point function on the disk, the first with two closed string vertex operators
and the second with one closed and one open string vertex operator, since
at leading order in $g_s N$ the absorption process results in the creation of a single massive
open string on the branes worldvolume. 
The possibility of having a closed-open transition induces a non-vanishing imaginary part for
the tree-level elastic scattering amplitude of the closed string.  
We start our
analysis by reviewing the elastic amplitudes, their imaginary part and their high energy limit. In this
paper we shall restrict our attention, for simplicity, to the bosonic string.  Generalization
to the superstring case is in principle straightforward and will be presented in \cite{clopunitary}.

The amplitudes describing the elastic scattering of a closed string are characterized by 
$t$-channel poles due to the exchange of closed strings and by $s$-channel poles due to intermediate physical open strings, which
will be analyzed in detail in the following Sections. For instance the simplest scattering amplitude, the one for the tachyon, reads~\footnote{The disk amplitudes for open and closed strings with Neumann boundary conditions were first computed in Ref.~\cite{Ademollo:1974fc}. Their extension to include Dirichlet boundary conditions and their application to the physics of the D$p$-brane were first
done in Refs. \cite{Klebanov:1995ni,Garousi:1996ad,Hashimoto:1996bf}.}  
\begin{equation}
  \label{2tachb}
  A_{TT} =  \frac{\kappa N T_p}{2} \frac{ \Gamma (-\alpha' s-1) \Gamma\left(-\frac{\alpha'}{4} t  -1\right)}{\Gamma \left( -\alpha' s -  \frac{\alpha'}{4} t -2\right)}~, \hspace{1cm} T_p = \frac{\sqrt{\pi}}{2^4}\si(2\pi\sqrt{\ap}\de)^{11-p} \ . 
\end{equation}
The normalization is given in terms of the gravitational coupling $\kappa^2 = 2^{-9} g_s^2 (2\pi)^{23} (\alpha')^{12}$, the number of branes $N$,
 and the brane tension $T_p$. The various quantities are taken for $d=26$.
The Mandelstam variables are defined in terms of the external momenta $p_1$ and $p_2$  of the two closed string tachyons
\ba
  t &=& -(p_1+p_2)^2 = -4 (E^2-M^2)\sin^2\frac{\theta}{2}~, \nb \\
s &=&-\frac{1}{4}(p_1 + D p_1)^2=-\frac{1}{4}(p_2 + D p_2)^2= E^2~,   \label{eq:s-t}
\ea
where $\theta$ is the angle between $\vec{p}_1$ and $-\vec{p}_2$, $M^2=-4/\alpha'$ and $D$ is the standard reflection matrix for a D$p$-brane, $D^\mu_{~\nu} = {\rm diag}(1,\ldots,1,-1,\ldots,-1)$ with 
the first $p+1$ eigenvalues equal to $1$ and the remaining $25-p$ equal to $-1$. Notice that we chose a reference frame where $(p_i + D p_i)$ has  no space-like components, so $\sqrt{s}$ is equal to the energy of the closed states. 
Using the relation
\be 
\frac{\Gamma(x)}{\Gamma(x+a)}= \sum_{n=0}^\infty \frac{ (-1)^n}{\Gamma(n+1)\Gamma(a-n)} 
\frac{1}{x+n} \  , 
\ee
with $a\not=0,1,\ldots$, we can write the amplitude in Eq.~\eqref{2tachb} in the  form\footnote{In deriving Eq. (\ref{eq:polesn}) we have used the $\Gamma$-function identity $ \Gamma (b) \Gamma (1-b) = \frac{\pi}{\sin \pi b}$.}
\begin{equation}
  \label{eq:polesn}
  A_{TT} =  \frac{\kappa N T_p}{2} 
\sum_{n=0}^\infty \frac{1}{n!} \, \frac{P_n\si( \frac{\ap t}{4}\de)}{-\ap s - 1 + n} 
 \ , \hspace{1cm}
P_n\si( \frac{\ap t}{4}\de) = \frac{\Gamma\left(\frac{\alpha'}{4} t + n + 2 \right) }{\Gamma\left(\frac{\alpha'}{4} t  +2\right) } \ ,
\end{equation}
that clearly displays the $s$-channel poles and their residues. 
There is a pole for each level of the open-string  spectrum, the residue being
a polynomial $P_n(y)$ of degree $n$ in $y=\frac{\ap t}{4}$.
The imaginary part of the amplitude follows from the usual $i\epsilon$ prescription and is obtained from~\eqref{eq:polesn} by substituting each pole $1/(-\alpha' s + m)$ with $\pi i \delta (\alpha' s -m)$ 
\begin{equation}
  \label{eq:polesnim}
 {\rm Im} A_{TT} =  \pi \frac{\kappa N T_p}{2}  
\sum_{n=0}^\infty \, \d\si(\ap s - n +1\de)  \frac{1}{n!} P_n\si( \frac{\ap t}{4}\de) \ . 
\end{equation}
In the Regge limit, $\ap s \gg 1$ with $\ap t$ fixed, the amplitude~\eqref{2tachb} behaves as\footnote{Here and in the following
we will use the symbol ${\cal A}$ for the Regge limit of an amplitude $A$.}
\begin{eqnarray}
A_{TT} \sim {\cal{A}}_{TT} =   \frac{ \kappa N T_p}{2} {\rm e}^{-i \pi \si ( 1 + \frac{\alpha'}{4}t\de)} \,
 \Gamma \si( -1 - \frac{\alpha'}{4}t\de)  \,  (\alpha' s)^{1+ \frac{\alpha'}{4}t  } \ . 
\label{Reggett}
\end{eqnarray}
Its real and imaginary part are easily evaluated
\begin{eqnarray}
&&{\rm Re} {\cal{A}}_{TT} =   \frac{ \kappa N T_p}{2} \cos \pi \left(1+\frac{\alpha'}{4}t\right) \Gamma \si( -1 - \frac{\alpha'}{4}t\de)  \,  
(\alpha' s)^{1+ \frac{\alpha'}{4}t  } \ ,  \nonumber \\
&& {\rm Im} {\cal{A}}_{TT} =  \pi  \frac{ \kappa N T_p}{2}\frac{ (\alpha' s)^{1+ \frac{\alpha'}{4}t}}{\Gamma \si(2 + \frac{\alpha'}{4}t \de) } \ . 
\label{reim}
\end{eqnarray}
It is interesting to derive the imaginary part in the Regge limit directly from 
the imaginary part of the full amplitude in Eq.~\eqref{eq:polesnim}. When $\ap s \gg 1$
the discrete distribution of poles can be approximated by a continuum, since
the relative separation between adjacent levels of the open string spectrum 
that are accessible at a given energy becomes of order $\frac{1}{\ap s}$.  
To study the contribution of the levels in a neighborhood of $\ap s $ let us set
\be
 n = \ap s x \ ,  \hspace{1cm} \sum_{n=0}^\infty  \d(\ap s-n+1)  \sim \int_0^\infty dx \, \d\si(1-x+\frac{1}{\ap s}\de) \ . \ee
In the limit $\ap s \gg 1$ we then reproduce\footnote{In deriving 
Eqs. (\ref{reim}) and (\ref{aaa}) we have used the large $a$ expansion $ 
\frac{ \Gamma (a + b)}{\Gamma (a+c) } \sim a^{b-c}$.}  Eq.~\eqref{reim}
\ba
 {\rm Im} {A}_{TT} &\sim&  \pi \frac{\kappa N T_p}{2}  
\int_0^\infty dx \, \d(1-x)   \frac{P_n\si( \frac{\ap t}{4}\de)}{\Gamma(n+1)} 
\sim   \pi \frac{\kappa N T_p}{2}  
\int_0^\infty dx \, \d(1-x) 
  \frac{(\ap s x)^{1+\frac{\ap t}{4}}}{\Gamma\si(2+\frac{\ap t}{4}\de)} \nb \\
&=&  \pi  \frac{ \kappa N T_p}{2}\frac{ (\alpha' s)^{1+ \frac{\alpha'}{4}t}}{\Gamma \si(2 + \frac{\alpha'}{4}t \de) }\ . 
\label{aaa}
\ea
In the following we will also be interested in scattering and absorption processes happening at fixed impact parameter. 
The imaginary part of the elastic disk amplitude
in impact parameter space is
\be
 {\rm Im}{\cal A}_{T T}(s,  b) = \int \frac{d^{24-p}q}{(2\pi)^{24-p}}
\, e^{-i \vec b \vec q} \,   {\rm Im}{\cal A}_{T T}(s, t) \sim  \pi  \frac{ \kappa N T_p}{2}   \frac{\ap s}{\si (\pi \ap \log \ap s \de)^{\frac{24-p}{2}}}
\, e^{-\frac{b^2}{\ap \log \ap s}} \label{imab} \ , \ee
where $\vec{q}$ is a $24-p$-dimensional vector satisfying $ -\vec{q}^2 \equiv t$.
In the last passage we approximated $\Gamma \si(2 + \frac{\alpha'}{4}t \de) \sim 1$. This approximation amounts
to neglecting at any given order in an expansion in
powers of $\frac{b^2}{\ap \log \ap s}$ terms that are suppressed by additional powers of $\log \ap s$. 
The imaginary part has a characteristic Gaussian dependence on the impact parameter, that indicates that in the Regge
limit the D$p$-branes behave as black disks with a radius growing like the square root of the logarithm of the energy,
$R \sim \sqrt{\ap \log \ap s}$. 

The elastic amplitude for the tachyon is extremely compact due to the fact that this state is a scalar with a simple
vertex operator 
\be {\cal V}_{T, \bar T} = \frac{\kappa}{2\pi}  e^{i p X} \ . \ee
The vertex operator for a generic closed string state $|\psi\ra$ can be written as follows
\be {\cal V}_{S, \bar S}  = \frac{\kappa}{2\pi} \, 
\e_{\m_1...\m_r} \bar \e_{\n_1...\n_s}
\, {\rm V}_S^{\m_1...\m_r} \, \bar{{\rm V}}_{\bar S}^{\n_1...\n_s} \ , \label{gvo} \ee
where $S$ and $\bar S$ are labels that identify the little group representations of the left and right
part of the closed state, $\e$, $\bar\e$ the corresponding polarization tensors and~\footnote{We write $X(z,\bar z) = X_L(z) + X_R(\bar z)$.}
\be {\rm V}_S = {\rm Pol}_S\si(\p^r X\de) \, e^{i p X_L} \ , \ee
with ${\rm Pol}_S$ a polynomial in the holomorphic derivatives of $X^\m$. 
Elastic and inelastic amplitudes for arbitrary states of the string spectrum become more and more complex as
the number of possible contractions between momenta and polarizations increases.

Their structure however drastically simplifies in the Regge limit and, as
discussed in~\cite{D'Appollonio:2013hja},  it is possible to give a general and explicit formula
in terms of the matrix elements of the phase of the eikonal operator~\cite{Amati:1987wq, Amati:1987uf, D'Appollonio:2010ae}. 
The Regge limit is characterized by a single spatial direction of large momentum that naturally leads to
a separation of the dynamics in a longitudinal and a transverse part. It is therefore convenient
to introduce a frame consisting of two light-like vectors $e^\pm$, whose spatial component
coincides with the direction of large momentum and that satisfy the following conditions
\begin{eqnarray}
e^+ \cdot e^+ = e^- \cdot e^- =0 \ , \hspace{1cm} e^+ \cdot e^- =1  \ , 
\label{e+-e}
\end{eqnarray}
together with $24$ transverse spatial vectors $e^i$, orthogonal to $e^\pm$. By convention we will choose
the large momentum along the spatial axis corresponding to the last coordinate 
 \begin{equation}
   e^+ = \frac{1}{\sqrt{2}} (-1,0,\ldots,0,1) \ , \hspace{1cm}
   e^-  = \frac{1}{\sqrt{2}} (1,0,\ldots,0,1) \ . \label{lc01}
 \end{equation}
When $\ap s \gg 1$ and $\ap t$ is kept fixed, the scattering
process is dominated by the exchange of the states of the leading Regge trajectory 
that can be represented by a single effective string state, the Reggeon~\cite{Ademollo:1989ag, Ademollo:1990sd, Brower:2006ea}
\be {\rm V}_{R} = \si (\sqrt{\frac{2}{\ap}} \, \frac{i \p X^+}{\sqrt{\ap}E} \de )^{1+\frac{\ap t}{4}} \, e^{- i q X_L}
\ , \hspace{1cm} X^+ = e^+ \cdot X \ .  \ee
In this limit the string-brane scattering amplitudes
factorize in the product of the three-point
coupling of the two external states to the Reggeon and the
one-point function of the Reggeon in the brane background, which coincides
with the Regge limit of the tachyon scattering amplitude, Eq.~\eqref{Reggett}. 
If the initial state is
$(S_1, \bar S_1)$ with polarization tensor $\e$, $\bar\e$ and
the final state $(S_2, \bar S_2)$ with polarization tensor $\z$, $\bar\z$ we can write
\be {\cal A}_{(S_1,\bar S_1), (S_2, \bar S_2)} =  {\cal A}_{TT} \, C_{S_1, S_2, R} 
\, \bar C_{\bar S_1, \bar S_2, \bar R} \label{reggeon} \ , \ee
where 
\be C_{S_1, S_2, R}  = \la {\rm V}_{S_1} {\rm V}_{S_2} {\rm V}_{R} \ra 
= \e_{\m_1...\m_r} \z_{\n_1...\n_s} \, T^{\m_1...\m_r; \n_1...\n_s}_{S_1, S_2, R} \ . \ee
The tensors $T_{S_1, S_2, R}$ are formed using the flat metric $\h^{\m\n}$, the momentum 
transferred $q^\m$ and the longitudinal polarization vector $v^\m$, with coefficients that depend
only on $t$ and the masses of the external states (see~\cite{D'Appollonio:2013hja} for details). 

The formula~\eqref{reggeon} can be simplified even further
if we describe the string spectrum using a basis of DDF operators rather than a basis of covariant
vertex operators \cite{D'Appollonio:2013hja}. In this way we only maintain manifest
invariance with respect to the transverse $SO(24)$ rotation group but it becomes straightforward to 
enumerate the physical states. Moreover the couplings of the external states to the Reggeon become elementary
and all the tree-level scattering amplitudes in the Regge limit can be represented as matrix elements of a very simple
operator $W(s, q)$, closely related to
the phase $\hat \d(s, q)$ of the eikonal operator. These simplifications occur if the null vector required by
the DDF construction is chosen proportional to $e^+$. In this way the amplitudes obtained using the
DDF operators coincide with the amplitudes given by the three-string vertex in the light-cone gauge
adapted to the kinematics of the Regge limit, as given in Eq.~\eqref{lc01}.

The operator $W$ that gives the tree-level scattering amplitudes is~\footnote{See \cite{D'Appollonio:2015gpa} for a derivation
of the eikonal operator for string-brane scattering in the bosonic string.}
\be W(s, q) =4 E \hat \d(s, q) =  {\cal A}_{TT}(s, t) \int_0^{2\pi} \frac{d\s}{2\pi} \, :e^{i q \hat{X}}: \ . \label{peb} \ee
The operators $X^i$ are the string position operators (without zero modes and at $\tau = 0$) 
\be \hat{X}^i(\s) = i \sqrt{\frac{\ap}{2}} \sum_{k \ne 0} \frac{1}{k} \si (A_k^i e^{-ik\s} +  \bar A_k^i e^{ik\s} \de ) \ , \ee
in a light-cone gauge with the spatial direction aligned with the direction of large momentum. 
The evaluation of the Regge limit of elastic or inelastic scattering amplitudes become straightforward
using the operator in Eq.~\eqref{peb}.
We illustrate this method with the elastic scattering of states in the massless and in the first massive
levels. 
A generic massless state (level $N_c=1$) can be written as
follows
\be |g_\e, \bar g_{\bar \e}\ra   = \e_{i}   \bar \e_{j} A_{-1}^i \bar A_{-1}^j |0\ra \ . \ee
The polarization tensor
\be \e_{ij} = \e_{i}   \bar \e_{j} \ , \ee
can be decomposed in a trace, symmetric traceless and antisymmetric part corresponding
respectively to the dilaton, graviton and Kalb-Ramond field. 
Denoting with $\e$ the polarization of the incoming state and with $\z$ the polarization of the outgoing state
we find
\be  {\cal A}_{gg} = \la g_\z, \bar g_{\bar \z} | \, W(s,q) \, |g_\e, \bar g_{\bar \e}\ra =  
\si ( \e \z - \frac{\ap}{2} \si (\e   q\de ) \si (\z   q\de )\de ) \si ( \bar \e   \bar \z - \frac{\ap}{2} \si (\bar \e  q\de ) \si (\bar \z   q\de )\de ) 
\, {\cal A}_{TT}\ . 
\ee
This is a general result. In the Regge limit all the scattering amplitudes are obtained by multiplying the tachyon amplitude
with a suitable polynomial in the polarizations and the momentum transferred. 
In impact parameter space the polynomial becomes a differential operator in $\p_{b_i}$ acting on $ {\cal A}_{TT}(s, b)$.
For instance
\be  {\cal A}_{gg}(s, \vec b) =
\si ( \e \z + \frac{\ap}{2} \si (\e   \p_b \de ) \si (\z    \p_b\de )\de )
 \si ( \bar \e   \bar \z + \frac{\ap}{2} \si (\bar \e   \p_b\de ) \si (\bar \z    \p_b\de )\de ) 
{\cal A}_{TT}(s, b)\ . 
\ee
The differential operator reflects the presence of higher derivative corrections to the gravitational couplings
in the bosonic string \cite{Camanho:2014apa, D'Appollonio:2015gpa}. As far as the imaginary part is
concerned the derivatives generate terms suppressed by additional powers of $\log( \ap s)$ and we
will neglect them for consistency with the approximations made in deriving Eq.~\eqref{imab}.
Therefore
\be  {\rm Im}{\cal A}_{gg}(s, \vec b) \sim \,  \e \z  \,  \bar \e   \bar \z \, \, {\rm Im}{\cal A}_{TT}(s, b)
 \ .
 \label{2.25} 
\ee
As an additional example of the general structure of the amplitudes let us consider the first massive level, 
that contains the $SO(25)$ representations in the tensor
product of two symmetric traceless tensors, one for the left and one for the
right half of the string. The $SO(25)$ symmetric traceless tensor decomposes into
a symmetric tensor and a vector of the transverse $SO(24)$ 
\be |H_\e \ra = \frac{1}{\sqrt{2}}\e_{ij}  A_{-1}^i A_{-1}^j   |0\ra\ , \hspace{1cm} 
 |L_\e \ra = \frac{1}{\sqrt{2}}\e_{i}  A_{-2}^i    |0\ra\ . \ee
The symmetric tensor can be further decomposed into a traceless and a trace part.
Let us discuss separately the elastic scattering of a closed state given by the product
of two  tensors $(H \otimes \bar H)$ and by the product
of  two vectors $(L \otimes \bar L )$. The remaining cases ($H \otimes \bar L $ and $L  \otimes \bar H$) can 
be analyzed in a similar way. For the state
\be |H_\e, \bar H_\e\ra = \frac{1}{2}\e_{ij} \bar \e_{kl} A_{-1}^i A_{-1}^j  \bar A_{-1}^k \bar A_{-1}^l |0\ra \ , \ee
we find 
\ba  {\cal A}_{H H} &=&  
\si ( \e_{ij} \z_{ij} - \frac{\ap}{2}    \e_{ii'} q^{i'} \z_{jj'} q^{j'} + \frac{\ap^2}{16} \e_{ij} q^i q^j \z_{i'j'} q^{i'} q^{j'}\de ) \nb \\
&& \si ( \bar \e_{kl}\bar  \z_{kl} - \frac{\ap}{2}   \bar \e_{kk'} q^{k'}\bar \z_{ll'} q^{l'}
+ \frac{\ap^2}{16} \bar \e_{kl} q^k q^l \bar \z_{k'l'} q^{k'} q^{l'}\de ) \, {\cal A}_{TT}\ .
\ea
For the state 
\be |L_\e, \bar L_\e\ra = \frac{1}{2}\e_{i} \bar \e_{j}  A_{-2}^i  \bar A_{-2}^j |0\ra \ , \ee
we find
\be  {\cal A}_{L L} =  
\si ( \e \z - \frac{\ap}{8} \si (\e q \de )\si (\z q\de )\de ) \si ( \bar \e \bar \z  - \frac{\ap}{8} \si (\bar \e q\de ) \si (\bar \z q\de )\de ) 
\, {\cal A}_{TT}\ .
\ee
A generic closed state $|\psi\ra$ at level $N_c$ is characterized by the collection of left and right modes that create it acting on the
vacuum 
\be A^{i_\a}_{-k_\a} \ , \hspace{1cm} \bar A^{i_\b}_{-\bar k_\b} \ , \hspace{1cm} 
\sum_\a k_\a = \sum_\b \bar k_\b = N_c \ , \ee
and by a collection of polarization vectors $\e_\a^{i_\a}$, $\bar\e_\b^{i_\b}$. 
To describe a state transforming in a specific irreducible representation of the transverse $SO(24)$, one simply
acts with the corresponding Young projector on the tensor product of the vector representations.
The elastic scattering amplitude 
\be {\cal A}_{\psi_\z \psi_\e} = {\cal A}_{TT}
 \int_0^{2\pi} \frac{d\s}{2\pi} \, \la \psi_\z | \, :e^{i q X}: \, |\psi_\e \ra  \ , \label{gea} \ee
can be evaluated by expanding the exponential and collecting the terms that give a non-vanishing matrix
element\footnote{In \cite{D'Appollonio:2013hja} it is shown in detail how to relate the matrix elements of the eikonal operator and the
covariant amplitudes.}. 
This will result in a polynomial where the polarization vectors of the incoming and outgoing states
are contracted among themselves or with the momentum transferred, like in the examples just discussed.
From~\eqref{gea} it is clear that at $t=0$ the elastic amplitude reduces to the contraction between the initial
and final polarization tensors
\be {\cal A}_{\psi_\z \psi_\e}(s, 0) = 
 \,\prod_\a \si ( \z_\a \e_\a \de )  \prod_\b \si (\bar \z_\b \bar \e_\b \de ) \, {\cal A}_{TT}(s, 0)
=  \pi  \frac{ \kappa N T_p}{2} \, \ap s  \,\prod_\a \si ( \z_\a \e_\a \de )  \prod_\b \si (\bar \z_\b \bar \e_\b \de ) \ . 
\label{imato}\ee
In impact parameter space the polynomial in the transferred momentum that multiplies the tachyon amplitude
${\cal A}_{T T}(s,  t) $
becomes a differential operator in $\p_{b_i}$ that acts on ${\cal A}_{T T}(s,  b) $.
Up to terms suppressed by additional powers of $\log \ap s$, the imaginary part 
is simply 
\be {\rm Im}{\cal A}_{\psi_\z \psi_\e}(s, b) = 
 \,\prod_\a \si ( \z_\a \e_\a \de )  \prod_\b \si (\bar \z_\b \bar \e_\b \de ) \, {\rm Im}{\cal A}_{TT}(s, b)\ . 
\label{imabo}\ee

\section{Absorption of closed strings by D-branes}
\label{abs}

In the background of a D$p$-brane a closed string can split turning itself into an open string. 
This process is described at tree level by a two-point function on the disk with one closed
and one open vertex operator. The closed-open correlation functions give the transition
amplitude to specific open string states while
the imaginary part of the elastic amplitude gives the total splitting probability.
We begin this Section by discussing the closed-open amplitude in the basis of the covariant
vertex operators. Since the covariant methods are not very suitable to 
study highly excited string states, we then introduce the light-cone closed-open vertex
that encodes all the correlation functions between arbitrary closed and open string states.  
In the rest of the paper we will develop methods that will
allow us to give a simple and precise description of the closed-open amplitudes in the
high energy limit.

\subsection{Closed-open amplitudes}

Consider the transition from a closed string at level $N_c$ 
with spatial momentum $\vec p_c$ in a direction orthogonal to the 
branes~\footnote{The components of the momenta are
given in the following order: $(t, vol, perp)$, where $t$ is the time direction,
$vol$ the $p$-dimensional branes volume, $perp$ the $(25-p)$-dimensional space transverse to the
branes.} 
\be p_c = (E, \vec 0_p, \vec p_c) \ , \hspace{1cm} \ap M^2 =  \ap (E^2 - \vec{p}{}^{\, 2}_c) =  4(N_c-1)\, \hspace{1cm}
N_c=0,1,2 \dots \ , 
\ee
to an open string at level $n$  with momentum
\be p_o = (-m, \vec 0_p, \vec 0_{25-p}) \ , \hspace{1cm} \ap m^2 = - \ap p_o^2 =  n-1 \, \hspace{1cm} n=0,1,2 \dots  \ . \ee
The transition is possible if
\be  \ap E^2 =  \ap m^2 = n - 1 \ , \hspace{1cm}  \ap \vec{p}{}^{\, 2}_c =  n - 4N_c + 3 \ge 0 \ . \ee
The lowest accessible level  is therefore $n=4N_c-3$. As the momentum of the closed string
increases it is possible to reach an arbitrarily high level of the open string spectrum.
For instance the tachyon can create
any open state, the massless states can create an open string 
with $n \ge 2$, closed states from the first massive  level can create an open string 
with $n=5$ and above. For the open vertex operators we use a notation similar to the one
introduced in Eq.~\eqref{gvo}
\be {\cal V}_\chi =  g_o \, \e_{\m_1...\m_r} \, {\rm V}_\chi^{\m_1...\m_r} \ , \ee
where $g_o = \sqrt{2\ap} g_{p+1}$, with $g_{p+1}$ the coupling constant of the gauge
theory living on the branes world-volume~\footnote{In terms of the string coupling constant $g_s$ and the Regge slope 
$\alpha'$ this constant is given by $g_{p+1}^{2} = 2 \pi g_s   
 (2\pi \sqrt{\alpha'})^{p-3}$. The Chan-Paton factors  are normalized as ${\rm Tr} (\lambda^i \lambda^j) =  \delta^{ij}$.}. 
When we have a stack of $N$ D$p$-branes, the open state carries Chan-Paton factors $\l^a$
that form a basis for the Lie algebra of $U(1) \times SU(N)$. Since the closed strings are singlets,
the open string state must belong to the $U(1)$ factor with $\l_s = \frac{1}{\sqrt{N}} \mathbb{I}_N$
and ${\rm tr}(\l_s) = \sqrt{N}$.
At tree level the absorption amplitude is given by the correlation function on the disk of two vertex
operators, one for the closed string state $\psi$  and one for the open string state $\chi$
\be B_{\psi, \chi}(p_c) = \b \, \la  {\rm V}_{S}  {\rm V}_{\bar S}   {\rm V}_{\chi} \ra_{D}   \ , \ee
where the normalization  $\b$ is given by 
\be  \b^2 = \frac{\kappa}{2\ap} N T_p \ . \label{beta} \ee 
The imaginary part of the two-point  elastic closed string amplitudes can be written as
\be {\rm Im}{\cal A}_{\psi,\psi}(p_1, p_2) = \pi \ap \sum_{\chi} B_{\psi, \chi}(p_1) B^*_{\psi, \chi}(p_2)
\ . \ee
The sum extends to all the open states $\chi$ at level $n$ and includes a sum over their polarizations.
Some explicit examples of closed-open amplitudes obtained by evaluating the correlation functions on the
disk can be found in appendix \ref{checks}. Since we are interested in the absorption of highly energetic
strings and therefore in very massive open states, the direct evaluation of correlation functions of vertex 
operators becomes soon prohibitively complex. 
We would like to have a simple method to derive the closed-open amplitudes and to characterize the
resulting open string state, akin to the formula for the closed-closed transition in Eq.~\eqref{gea}.
We will develop this method in the following Sections 
by studying the high-energy limit of the light-cone closed-open string vertex, to which we now turn.

\subsection{The closed-open string vertex}

The closed-open light-cone vertex describes the transition from an arbitrary  closed string state to an arbitrary  open state living on a 
D$p$-brane.
The closed-open vertex for the bosonic string for the case of purely Neumann boundary conditions was first discussed in 
\cite{Cremmer:1973ig, Clavelli:1973uk}
and then generalized to include ghosts \cite{Shapiro:1987gq} and purely Dirichlet boundary conditions \cite{Green:1994ix}.
The extension to the superstring in the Green-Schwarz formalism was considered in \cite{Green:1983hw}.
Here we give the explicit form of the vertex, which we derived using 
the DDF operators as in \cite{Ademollo:1974kz}, for a generic D$p$-brane background. 

The light-cone is determined by two light-like vectors  $e^{\pm}$ satisfying the following conditions
\begin{eqnarray}
e^+ \cdot e^+ = e^- \cdot e^- =0 \ , \hspace{1cm} e^+ \cdot e^- =1  \ . 
\label{e+-}
\end{eqnarray}
There are three inequivalent choices for the spatial direction of the light-cone: along the direction of the large
momentum carried by the closed string, along the volume of the D$p$-branes 
or transverse to both the branes and the direction of large momentum.  We will discuss explicitly
the first two choices. As we will see, the open state created at high energy has an extremely simple description if the light-cone
is chosen along the direction of large momentum, as we did in the previous Section in our discussion
of the Regge limit of the closed-closed scattering amplitudes and their relation to the eikonal operator.

Having chosen $e^{\pm}$ the light-cone vertex  is given  by
\begin{equation}
|V_{B}\rangle = \b \, \exp\left[    \sum_{r, s = 1}^{3} \sum_{k, l=1}^\infty
\frac{1}{2} A_{-k} ^{r, i}N^{rs}_{k l} A^{s, i}_{-l}  +  \sum_{r=1}^{3} \sum_{k=1}^\infty P^i 
N_{k}^{r} A_{-k}^{r, i} \right] \prod_{r=1}^{3}\left(  | 0\rangle^{(r)}
\right) \ , 
\label{3VB}
\end{equation}
where 
\begin{equation}
P_i \equiv \sqrt{2\alpha'}\left[ \alpha_r p^{(r+1)}_{i} - \alpha_{r+1} p^{(r)}_{i}\right]~.
\label{Pix}
\end{equation}
Let us explain our notation. 
The normalization
constant $\b$ is given in Eq.~\eqref{beta} and it is related to the normalization of the closed 2-point function on the disk.
The index $i$ runs along the $d-2$ directions orthogonal to both $e^{\pm}$ ($d=26$ for the bosonic string) and 
the quantities labelled by $r, s = 1, 2$ refer to the left/right parts of the closed string, while those labelled by $r, s = 3$ refer to the open state.  
In particular
\be p^{(1)} = \frac{p_c}{2} \ , \hspace{1cm} p^{(2)} = \frac{D p_c}{2} \ , \hspace{1cm}
p^{(3)} = p_o \ , \label{pi89} \ee
where $p_c$ is the momentum of the closed string~\footnote{The components of the momenta are
given in the following order: $(t, vol, perp, z)$, where $t$ is the time direction,
$vol$ the $p$-dimensional branes volume, $perp$ the $(24-p)$-dimensional space transverse to the
branes and to the direction $z$  of the large momentum $p$.} 
\be p_c = (E, \vec 0_p, \vec p_t, p) \ , \hspace{1cm} \ap M^2 = - \ap p_c^2 =  4(N_c-1) \ , \ee
and $p_o$ the momentum of the open string
\be p_o = (-m, \vec 0_p, \vec 0_{24-p}, 0) \ , \hspace{1cm} \ap m^2 = - \ap p_o^2 =  n-1 \ . 
\ee
Finally, the harmonic oscillators satisfy the following commutation relations
\begin{eqnarray}
[ A_{k}^{r,i} , A_{h}^{s,j} ] =  k\, \delta_{h+k, 0}\, \delta^{rs}\, \delta^{ij}  \ . 
\label{commurela}
\end{eqnarray}
When considering the high energy limit we will choose, as in the previous Section, 
the large momentum along the spatial axis corresponding to the
last coordinate, i.e.  $p^2 \gg \vec p_t^2$. 
We will use capital letters for the light-cone modes of the closed string and small case letters for the light-cone
modes of the open string
\be A^{1,i}_{k} = A_k^i \ , \hspace{1cm} A^{2,i}_{k} = (D \bar A^i)_k \ , \hspace{1cm}
A^{3,i}_{k} = a^i_k \ . \label{DA} \ee
Finally the Neumann matrices $N$ in the vertex are 
\begin{align}
\label{eq:Nmn}
N_{k}^{r} & = - \frac{1}{k \alpha_{r+1}}\left(  \begin{array}{c} - k \frac{\alpha_{r+1}}{\alpha_r} \\ k \end{array} \right) =
 \frac{1}{\alpha_r k!} \frac{ \Gamma  \left(   - k 
\frac{\alpha_{r+1}}{\alpha_r} \right)  }{ \Gamma  \left(   - k 
\frac{\alpha_{r+1}}{\alpha_r} +1 - k \right) } \ , \\
N^{rs}_{k l} &  =  - \frac{k l \alpha_1 \alpha_2 \alpha_3}{k \alpha_s + l \alpha_r} N^{r}_{k} N_{l}^{s} =  - \frac{\alpha_1 \alpha_2 \alpha_3}{k\alpha_s + l\alpha_r} \frac{1}{\alpha_{r+1} \alpha_{s+1}} \left( \begin{array}{c} - k \frac{\alpha_{r+1}}{\alpha_r} \\k \end{array} \right)
\left( \begin{array}{c} - l \frac{\alpha_{s+1}}{\alpha_s}  \\ l\end{array} \right) \ , 
\nonumber
\end{align}
while the quantities $\alpha_r$ are given by
\begin{eqnarray}
\alpha_r = 2 \sqrt{2\alpha'} (e^+ p^{(r)})\ , \hspace{1cm} r=1,2,3\ , 
\label{alphaq}
\end{eqnarray}
with 
\begin{eqnarray}
\alpha_1 + \alpha_2 + \alpha_3 =0\,\ .
\label{sumuptozero}
\end{eqnarray}
The momenta $P_i$ and the  light-cone components $\a_i$  depend on the choice of gauge.
We give below the explicit expressions  for 
two inequivalent choices of the spatial components of the light-cone vectors $e^{\pm}$: along the direction
of the large momentum carried by the closed string and  along the worldvolume of the D$p$-branes.

The first choice can be made for every D$p$-brane background. The light-cone vectors are
 \begin{equation}
   \label{eq:lico}
  e^+  = \frac{1}{\sqrt{2}} (-1,0,\ldots,0,1) \ , \hspace{0.6cm}
   e^-  = \frac{1}{\sqrt{2}} (1,0,\ldots,0,1) \ ,
 \end{equation}
and the $\alpha_r$  are equal to
\begin{align}
\alpha_1 &  = \sqrt{\alpha'} \left( E + p\right) = \sqrt{n-1}  + \sqrt{n-1 - 4\w } \ , 
\nonumber \\ 
\alpha_2 &   = \sqrt{\alpha'} \left( E - p\right) = \sqrt{n-1}  - \sqrt{n-1 - 4\w } \ , 
\label{alpha1alpha2} \\ \nonumber 
\alpha_3 &  = -2 \sqrt{\alpha'} E  = -2 \sqrt{n-1} \ , 
\end{align}
where
\be \w = \frac{\ap}{4} (M^2+ \vec p_t{}^2)  = N_c-1+ \frac{\ap \vec p_t{}^2}{4}  \ . \label{mu} \ee
In this light-cone gauge the index $i$ runs over the directions  transverse to the branes
and to the direction of large momentum, i.e. $i = p+1, ..., 24$.
Using Eq.~\eqref{pi89} we find
\begin{equation}
  \label{eq:pi}
  P_i = \alpha_3 \, \sqrt{\frac{\ap}{2}} \, p_{t, i}\, . 
\end{equation}
Note that the argument of the Gamma functions in the Neumann coefficients depends on the value of the energy
and of the transverse momentum.

The second gauge choice is possible only for D$p$-branes with $p \ge 1$. The light-cone vectors are
 \begin{equation}
   \label{eq:lico2}
  e^+  = \frac{1}{\sqrt{2}} (-1,1,\ldots,0,0)\ , \hspace{0.6cm}
   e^- = \frac{1}{\sqrt{2}} (1,1,\ldots,0,0) \ .
 \end{equation}
We then find
\begin{eqnarray}
\alpha_1 = \alpha_2=\sqrt{\alpha'} E\ , \hspace{0.6cm} \alpha_3 = - 2\sqrt{\alpha'} E \ , 
\label{alphas}
\end{eqnarray}
and 
\begin{eqnarray}
P_i =  \alpha_3 \,  \sqrt{\frac{\ap}{2}} \,p_{c, i}  \ ,
\label{Pi}
\end{eqnarray}
where $p_{c, i}$ are the components of  the momentum of the closed string along the transverse directions. 
In this light-cone gauge the index $i$ runs over all directions transverse to the brane, $i = p+1,..., 25$.
In this case the argument of the Gamma functions in the Neumann coefficients does not depend on the value of the energy
and of the transverse momentum.

In the rest of this paper we will work mostly in the first gauge, where we will find a very simple representation for
the massive open string.  
Before continuing, there is a subtlety related to this gauge choice for the open string  that deserves some explanations. 
The gauge choice $X^+ = 2 \ap p^+ \t$ together with the Virasoro constraints requires that the 
two open string coordinates chosen to form the light-cone satisfy Neumann boundary conditions. 
This seems incompatible with defining the coordinates $X^\pm$ for the open string states as the combination of the time 
coordinate $X^0$, which satisfies Neumann boundary conditions, and 
the direction of collision $Z$, which satisfies Dirichlet boundary conditions.
For instance, for any given $X^0$ and $Z$ the reparametrization of the open worldsheet required to set $X^+ = 2 \ap p^+ \t$
does not leave the worldsheet boundaries fixed. 

The way around this problem is to define a modified light-cone gauge using for the open string instead  of the coordinate $Z$
a coordinate $\tilde Z$ given by~\footnote{We write $X(\t, \s) = X_L(\s^+) + X_R(\s^-)$.}   
\be Z(\t,\s ) = z(\s^+) - z(\s^-) \ \ \ \ \  \mapsto \ \ \ \ \
  \tilde Z(\t,\s ) = z(\s^+) + z(\s^-) \ , \ee 
where $\s^\pm = \t \pm \s$. 
The coordinate $\tilde Z$ satisfies Neumann boundary conditions and there is no subtlety in fixing the light-cone gauge
$\tilde X^+ = 2 \ap \tilde p^+ \t$, where  $\tilde X^\pm = \frac{1}{\sqrt{2}}(X^0 \pm \tilde Z)$ and $\sqrt{2} \tilde p^+
= p^0_o$. 
Since the massive open strings do not have a momentum zero-mode, their description in terms of the coordinate $Z$ or
the coordinate $\tilde Z$ is equivalent~\footnote{In terms of the original coordinates
this gauge choice sets $  X^+_L(\sigma^+) =  \alpha' p^+_L \sigma^+$ and
  $X^-_R(\sigma^-) =  \alpha' p^-_R \sigma^-$, with $\sqrt{2}p_L^+ = \sqrt{2}p_R^- = p^0_o$. }.

In Eq.~\eqref{3VB} we wrote the vertex as a ket in the closed-open Hilbert space. Equivalently, we could represent it 
as an operator mapping the closed string Hilbert space to the open string Hilbert space, simply by replacing all
the closed string creation operators with annihilation operators. 
Given a closed state $|\psi\ra$ at level $N_c$ and with energy $\ap E^2 = n - 1$, 
the closed-open vertex gives its couplings to all the open string states $|\chi\ra$ at level $n$
\be B_{\psi \chi} = \la \chi| V |\psi \ra \ . \ee
The imaginary part of the elastic disk amplitude with two external closed string states 
$|\psi_1 \ra$ and $|\psi_2 \ra$ can be computed
combining two closed-open vertices. 
At finite energy it is convenient to choose in our first light-cone gauge
a brick-wall frame where the momenta of the incoming and outgoing closed strings are given by
\begin{eqnarray}
p_1 = \si(E, \vec 0_p, \frac{\vec q}{2} , p\de) \ , \hspace{1cm}
p_2 = \si(-E, \vec 0_p, \frac{\vec q}{2} , -p\de)\ , 
\label{popenpclosed}
\end{eqnarray}
so that $t = -(p_1+p_2)^2 = - \vec q {}^{\, 2}$. The 
angle $\theta$ between $\vec{p}_1$ and $-\vec{p}_2$ is given by
\begin{equation}
  \label{eq:betatheta}
\frac{\vec q^{\, 2}}{4}  = \si(E^2-M^2\de) \sin^2 \frac{\theta}{2} \ , \hspace{1cm} p^2 = \ \si(E^2-M^2\de) \cos^2 \frac{\theta}{2}\ .
\end{equation}
We can then write 
\be {\rm Im}A_{\psi \psi}(s, t) = \pi \ap
 \la  \psi_2 | V_{ \vec q/2 }^\dagger  V_{ \vec q/2} |\psi_1\ra
\ , \hspace{1cm} \ap s = n - 1 \ , \hspace{0.4cm} \ap t = - \ap \vec q {}^{\, 2} \ . \label{img1} \ee
Here and in the following, we show explicitly the dependence of the closed-open vertex 
on the transverse momentum carried by the closed state on which it acts.
We will use this choice of frame to check that the closed-open vertex correctly reproduces the imaginary
part of the elastic scattering of the tachyon for the first  massive level of the spectrum in Appendix
\ref{checks}. 
In the Regge limit it is sometimes more convenient to choose
\begin{eqnarray}
p_1 = \si(E,  \vec 0_p,  \vec q , \sqrt{p^2 - \vec q{}^{2}} \de) \ , \hspace{1cm}
p_2 = \si(-E,  \vec 0_p,  \vec 0_{24-p} , - p  \de)\ . 
\label{popenpclosed2}
\end{eqnarray}
In this frame the relation $t = -(p_1+p_2)^2 \sim - \vec q {}^{\, 2}$ remains valid up to terms of order $s^{-1}$, which
are negligible in the high energy limit if we are only interested in the leading order.

\section{The high-energy limit of the closed-open vertex}
\label{henc}

In the previous Section we summarized the form of the vertex that gives the transition amplitude
between an arbitrary closed string state and an arbitrary open string state. We now discuss how
the structure of this operator simplifies when the absorbed string is ultrarelativistic, $\ap s \gg 1$ and $M \ll E$.
In this Section and in the rest of the paper
we will work in the light-cone gauge parallel to the collision axis, or more precisely to the direction of large momentum,
since it is in this gauge that
we will be able to give a very explicit description of the massive open state created on the brane worldvolume.
As a check of our construction we will verify that starting from the high-energy limit of the closed-open vertex  
we can reproduce the 
imaginary part  of the elastic amplitude, which is due to the creation of on-shell states in the $s$-channel. 
This analysis complements the analysis performed in~\cite{D'Appollonio:2013hja} where
we studied the Regge limit of the three-closed string light-cone vertex, which gives the
$t$-channel decomposition of the elastic amplitude.

Consider a closed state $|\psi\ra$ at level $N_c$, with energy $E$ such that 
\be \ap E^2 = n - 1 \ , \ee
and carrying a momentum $\vec p_t$ in the transverse directions, $p^2_t \ll E^2$.
The non-trivial part of the closed-open vertex in Eq.~\eqref{3VB} when acting on this state can be written as
\be V_{\vec p_t} |\psi\ra = \b \, {\cal P}_n \,  e^{Z_o+Z_{c,\psi}} \, |\psi\ra \ , \label{vcog}\ee
where $Z_o$  contains only the open string modes (i.e. the Neumann coefficients $N^{33}_{k l}$ and $N^3_{k}$)
and $Z_{c,\psi}$ contains all the remaining terms which have both open and closed string modes, 
the latter restricted to those that can have a non-vanishing
contraction with the modes used to define the closed state
The operator ${\cal P}_n$ is the projector on the level $n$ of the open string Hilbert space
and enforces energy conservation. 
We shall derive the asymptotic behaviour of the vertex~\eqref{vcog} in the large $n$ limit
\be V_{\vec p_t} \sim {\cal V}_{\vec p_t} \ , \hspace{1cm} n \gg 1 \ . \ee
The asymptotic vertex gives the couplings between a highly energetic closed string $|\psi\ra$ 
and a very massive open string $|\chi\ra$
\be {\cal B}_{\psi \chi} = \la \chi| {\cal V}_{ \vec p_t}|\psi\ra \ . \ee
The imaginary part of the elastic disk amplitude in the Regge limit can be computed
combining two closed-open vertices
\be {\rm Im}{\cal A}_{\psi \psi}(s, t) = \pi \ap
 \la  \psi | {\cal V}_{ \vec 0}^\dagger  {\cal V}_{ \vec q} |\psi\ra
\ , \hspace{1cm} \ap s = n - 1 \ , \hspace{0.4cm} \ap t = - \vec q^{\, 2} \ . \label{img} \ee
We will also be interested in the closed-open couplings and in the imaginary part of the elastic disk amplitude
in impact parameter space 
\ba {\cal B}_{\psi \chi}(\vec b) &=& \int \frac{d^{24-p}q}{(2\pi)^{24-p}}
\, e^{-i \vec b \vec q} \, \la \chi|  {\cal V}_{\vec q} |\psi\ra \equiv \la\chi| \widetilde{{\cal V}}_{\vec b} |\psi\ra \ , \ea
\ba
 {\rm Im}{\cal A}_{\psi \psi}(s, \vec b) &=& \pi \ap \,  \int \frac{d^{24-p}q}{(2\pi)^{24-p}}
\, e^{-i \vec b \vec q} \, 
 \la  \psi | {\cal V}_{ 0}^\dagger  {\cal V}_{\vec q} |\psi\ra \equiv 
\pi \ap  \la  \psi | {\cal V}_{ 0}^\dagger  \widetilde{{\cal V}}_{\vec b} |\psi\ra \,  \ , \label{imgb} \ea
where we introduced the vertex in impact parameter space
\be  \widetilde{{\cal V}}_{\vec b} = \int \frac{d^{24-p}q}{(2\pi)^{24-p}}
\, e^{-i \vec b \vec q} \, {\cal V}_{\vec q} \ . \ee
As we shall see, at high energy the Fourier transforms in the previous expressions are dominated by the region of small transverse momenta
and therefore we will find a simple relation between the vertex in impact parameter space and the vertex at zero transverse
momentum.

In order to derive the high energy limit of the vertex, we must determine the asymptotic behaviour
of  the Neumann coefficients.
At high energy the ratios of the light-cone components $p^+$ of the momenta 
 of the three strings that appear in Eq.~\eqref{eq:Nmn} behave as follows
\ba   \frac{\a_3}{\a_2} &=& - \frac{2m}{E-p} \sim - \frac{n}{\w} \ , \nb \\
  \frac{\a_2}{\a_1} &=& \frac{E-p}{E+p}  \sim   \frac{\w}{n} \ , \nb \\
 \frac{\a_1}{\a_3} &=& - \frac{E+p}{2m} \sim -1 + \frac{\w}{n} \ ,
\ea
where the $\a_i$ are defined in Eq.~\eqref{alpha1alpha2} and $\w$ is defined  in Eq.~\eqref{mu}.
Using these expressions we can analyze the high energy limit of the Neumann coefficients
$N^{rs}_{k l}$ and $N^r_{k} \vec P$. 
When  $n$ tends to infinity, the coefficients quadratic in the open modes behave like 
\ba  N^{33}_{k l} &\sim& - \frac{\w}{n}
 \frac{ 1}{ k+ l} \, \frac{k^{-\w \frac{k}{n}}}{\Gamma\si(1-\w \frac{k}{n}\de)} \, \frac{l^{-\w \frac{l}{n}}}{\Gamma\si(1-\w \frac{l}{n}\de)}\ . \ea
If in the large $n$ limit the ratios $k/n$ and $l/n$ tend to zero this is 
\be   N^{33}_{k l} \sim - \frac{\w}{n}
 \frac{ 1}{ k+ l} \ . \ee
On the other hand, if the open modes are of order $n$ setting
\be k = n x \ , \hspace{1cm} l = n y \ , \label{cont}\ee
we find 
\ba
N^{33}_{k l} &\sim& -  \, n^{-\w(x+y)-2} \,
 \frac{x^{-\w x}y^{-\w y}}{\Gamma(1-\w x)
\Gamma(1-\w y)} \frac{ \w}{ x+y} \  . \label{n33he} \ea
The coefficients quadratic in the closed modes have the following behaviour 
\ba  
N^{11}_{k l} &\sim&  \frac{\w}{n} \, \frac{  (-1)^{k+l}}{ k+ l}  \ , \nb \\
N^{22}_{k l} &\sim& \si(\frac{n}{\w}\de)^{k+l}  \frac{ k^{k}}{k! } \,
  \frac{ l^{l}}{ l!}
 \frac{1}{ k+ l} \ , \nb \\
N^{12}_{k l}  &\sim&  (-1)^{k+1} \si(\frac{n}{\w}\de)^{l-1}
\frac{ l^{l-1}}{l!} 
\ . \ea
The left and right closed string modes $k$, $l$ are always much smaller than $n$. 
Note that the coefficients $N^{22}_{kl}$ scale like a positive power of $n$.
The mixed coefficients with one closed and one open mode behave generically like
\ba 
N^{13}_{k l } &\sim& \frac{(-1)^{k}}{l-k}  \frac{\w}{n } 
\, \frac{l^{-\w \frac{l}{n}}}{\Gamma\si(1-\w \frac{l}{n}\de)}
 \ ,  \nb \\
N^{23}_{k l} &\sim&  \frac{1}{k-\w\frac{l}{n}} \, \frac{ k^{k}}{k!}\,
  \si(\frac{n}{\w}\de)^{k-1} \, \frac{l^{-\w \frac{l}{n}}}{\Gamma\si(1-\w \frac{l}{n}\de)}
  \ .
\ea
The coefficients $N^{13}_{kl}$ possess an interesting feature. They
are enhanced by an additional power of $n$ when the open  and the closed indices coincide
\be 
N^{13}_{k k } \sim \,  - \frac{ (-1)^{k} }{k} \frac{n}{\omega k} \frac{\sin\left(\pi \omega \frac{k}{n}\right)}{\pi} \sim 
\, - \frac{ (-1)^{k} }{k}
\ , \label{enhance}\ee
where the last approximation holds since $k\ll n$.
This feature will lead to a simple formula for the absorption of a generic closed string state. 
Finally the coefficients linear in the string modes behave as follows
\ba N^1_{k} \, \vec P &\sim&  \frac{(-1)^{k}}{k} \, \sqrt{\frac{\ap}{2}} \vec p_t \ , \nb \\
N^2_{k}  \, \vec P &\sim& -  \frac{k^k}{k!} \, \si(\frac{n}{\w}\de)^{k}  \frac{1}{k} \, \sqrt{\frac{\ap}{2}} \vec p_t  \ , \nb \\
N^3_{k} \, \vec P  &\sim&  \,   \frac{k^{-\w \frac{k}{n}}}{\Gamma\si(1-\w \frac{k}{n}\de)}
  \frac{1}{ k}\, \sqrt{\frac{\ap}{2}} \vec p_t  \ . \ea

\section{Closed-open transitions at high energy}
\label{heco}

We are now ready to derive the form of the open state created in the absorption process. 
For clarity, we will analyze separately the absorption of a tachyon,
of a massless state and of a state belonging to the first massive level. We will then describe the open state
created by the absorption of an arbitrary closed state. 

We will discuss first the case in which the closed state carries zero transverse momentum
and show how to reconstruct the discontinuity 
of the elastic amplitude at $t=0$. We will then include a transverse momentum  $\vec p_t$ (with $p_t^2 \ll E^2$)
and derive the form of the open state created at fixed impact parameter $b$, showing how 
to reconstruct the discontinuity of the elastic amplitude in impact parameter space
for arbitrary values of $b$. 

\subsection{Tachyon}

We begin for simplicity with the vertex that describes the transition of a closed string tachyon with $\vec p_t = 0$
to an open string state with mass $\ap m^2 = n - 1$
\ba {\cal V}_{\vec 0}|0\ra &=& \b \, {\cal P}_n \, e^{\frac{1}{2} \sum_{k,l} N^{33}_{k l} a^i_{-k} a^i_{-l}}|0\ra \ , \label{v1} \ea
where for large $n$
\ba
N^{33}_{k l} &\sim&   \, n^{x+y-2} \,
 \frac{x^{ x}y^{ y}}{\Gamma(1 + x)
\Gamma(1 + y)} \frac{1}{ x+y} \ , \hspace{1cm} k = n x \ , \hspace{0.3cm} l = n y  , \label{n33t} \ea
which follows from Eq.~\eqref{n33he} with $\w = - 1$.
Expanding the exponential in Eq.~\eqref{v1}, we obtain the following series representations for the open state
\be  {\cal V}_{ \vec 0}|0\ra = \b \, \sum_{Q=0}^\infty \,  \, \frac{1}{Q! 2^Q} \, {\cal P}_n \, 
\prod_{\a=1}^Q \sum_{k_\a,l_\a} N^{33}_{k_\a l_\a} a^{i_\a}_{-k_\a} a^{i_\a}_{-l_\a}
|0\ra \ , \label{heco1} \ee
and for the imaginary part of the elastic amplitude
\be  {\rm Im}{\cal A}_{TT} =  \pi \ap \b^2  \,  \sum_{Q=0}^\infty \,  \, \frac{1}{(Q!)^2 2^{2Q}} 
 \sum_{\{k_\a,l_\a\} } \prod_{\a=1}^Q \si ( N^{33}_{k_\a l_\a}\de)^2 
 \la 0 |  \prod_{\a=1}^Q a^{i_\a}_{k_\a} a^{i_\a}_{l_\a}  \prod_{\a=1}^Q a^{i_\a}_{-k_\a} a^{i_\a}_{-l_\a}|0\ra 
\ , \label{Qst} \ee
where the integers $k_\a, l_\a$ satisfy the constraint $\sum_{\a=1}^Q(k_\a+l_\a) = n$.
It is remarkable that at high energy and in the  gauge aligned to the large momentum, it is sufficient to keep only the first few terms
in the previous series in order to get a very accurate representation of the state. 

Let us evaluate the first term in the series in Eq.~\eqref{Qst},
which approximates the open state with the state created by the action of just one couple of open modes
$a^{i}_{-k} a^{i}_{-l}$
\ba {\rm Im}{\cal A}_{TT}  &\sim& 
\pi \ap \b^2 \sum_{k, l\atop k+l=n} \frac{1}{4} \si( N^{33}_{k l} \de)^2  \la 0 |  a^i_{k} a^i_{l}  a^j_{-k} a^j_{-l}|0\ra 
=  \pi \ap \b^2 \sum_{k, l\atop k+l=n} 12 \si( N^{33}_{k l} \de)^2 k l   \ , 
\ea
where we used the fact that the indices $i,j$ run from $1$ to  $24$.
At high energy we can evaluate the previous sum by approximating it with an integral, using Eq.~\eqref{n33t}
\ba 
{\rm Im}{\cal A}_{TT}  &\sim&  12  \, \pi \ap \b^2  \, n \, \int_0^1 dx \int_0^1 dy \d(x+y-1) 
\frac{x^{2x+1}y^{2y+1}}{\Gamma^2(1+x)\Gamma^2(1+y)} \ . 
\label{5.6}
\ea 
Evaluating the integral we find that this approximation to the complete open state already accounts for
$0.93$ of the total imaginary part
\be  {\rm Im}{\cal A}_{TT}  \sim  0.929  \, \pi \ap \b^2 \, n \ . \ee
Note that since in this case the product of the  Neumann coefficients $\si( N^{33}_{k l} \de)^2 k l$ give a contribution
of order one, the  power
of $n$ required by the discontinuity of the elastic amplitude is entirely due to the multiplicity of  the possible
states (the number of partitions of $n$ into two integers).
In general for any given  term of the series in Eq.~\eqref{Qst} 
containing $Q$ couples of oscillators the overall power of $n$ results from a power of $n^{2-2Q}$ 
from the product of the Neumann coefficients and a power of $n^{2Q-1}$ from the sum over all possible states. 

Consider for instance  the second term in the series. There are now eight modes in the matrix
element and  two classes of inequivalent
contractions among them, giving respectively a double trace and a single trace in the transverse indices.
In the first case we have 
\be 72 \,    \int dx_1 dy_1 dx_2 dy_2 \prod_{i=1}^2 \frac{x_i^{2x_i+1}y_i^{2y_i+1}}{\Gamma^2(x_i+1)\Gamma^2(y_i+1)} 
\frac{\d\si(x_1+y_1+x_2+y_2 -1\de)}{(x_1+y_1)^2(x_2+y_2)^2}
 \ , \ee
and in the second
\be 6 \,   \int dx_1 dy_1 dx_2 dy_2 \prod_{i=1}^2 \ \frac{x_i^{2x_i+1}y_i^{2y_i+1}}{\Gamma^2(x_i+1)\Gamma^2(y_i+1)} 
\frac{\d\si(x_1+y_1+x_2+y_2 -1\de)}{(x_1+y_1)(y_1+x_2)(x_2+y_2)(y_2+x_1)}
 \ , \ee
where in both cases we omitted a factor $\pi \ap \b^2 n$. 
Adding the contribution of  these two terms to the leading contribution we obtain
\be   
{\rm Im}{\cal A}_{TT} \sim  0.998  \, \pi \ap \b^2  \, n\ . 
\label{5.10}
\ee
Therefore only $0.2 \%$ of the full forward imaginary part is left to terms with six or
more oscillators. This is strong evidence that the series converges rapidly and that
\be \la 0| e^{Z_0^\dagger}{\cal P}_n e^{Z_0}|0\ra = n \ , \label{conjsum} \ee
as required by Eq.~\eqref{img}. 
 
Let us now consider  the absorption amplitude and  the imaginary part of the disk at fixed impact parameter.
We will find that the open state is closely related to the one created by the tachyon when $\vec p_t = 0$. 
We start from the vertex with a non-vanishing transverse momentum $\vec q$ and evaluate
its Fourier transform. 
For clarity, in the following we will display explicitly the dependence of the Neumann coefficients
on the transverse momentum $\vec q$, writing $N^{rs}_{kl}\si(\vec q\de)$ and $N^r_k(\vec q)$.

When the transverse momentum is non zero, the vertex becomes
\ba  {\cal V}_{\vec q}|0\ra &=& \b \, {\cal P}_n  \, e^{\frac{1}{2} \sum_{k,l} N^{33}_{k l}(\vec q) a^i_{-k} a^i_{-l} +\sum_k 
N^3_{k}(\vec q) \, P^i a^i_{-k}}|0\ra \ , \label{vtq} \ea
where 
\ba
N^{33}_{k l}(\vec q) &\sim&   \, 
 \frac{k^{(1-\l) \frac{k}{n}} l^{(1-\l) \frac{l}{n}}}{\Gamma\si(1+(1-\l)\frac{k}{n}\de)
\Gamma\si(1+(1-\l) \frac{l}{n}\de)} \frac{(1-\l)}{n(k+l)} \  , \hspace{1cm} \l = \frac{\ap \vec q^{\, 2}}{4} \ , \nb \\
N^3_{k}(\vec q) \, \vec P  &\sim&  \,  \frac{k^{(1-\l)\frac{k}{n}}}{\Gamma\si(1+(1-\l)\frac{k}{n}\de)}\, \frac{1}{k}
\,\sqrt{\frac{\ap}{2}}\vec q  \  . \label{n33tq} \ea
Expanding  the exponential in Eq.~\eqref{vtq}, it is easy to see that the 
linear and the quadratic Neumann coefficients give contributions of the same order in the energy. 
It is however possible to find a very simple form for the closed-open vertex in impact parameter space. 
The main observation is that the essential dependence of the coefficients $N^{33}_{k l}(\vec q)$ and $N^3_{k}(\vec q)$ 
on the transverse momentum is in the
factors $k^{-\l \frac{k}{n}}$ or, with $k = nx$, 
\be (nx)^{-\l x} = e^{- \frac{\ap \vec q^{\, 2}}{4} x \log (nx)} \sim e^{- \frac{\ap \vec q^2}{4} x \log n} \ , \label{expq}\ee
since $x \log x \ll x \log n$. 
In order to evaluate
\be  \widetilde{{\cal V}}_{ \vec b}|0\ra =  \int \frac{d^{24-p}q}{(2\pi)^{24-p}}
\, e^{- i \vec b \vec q} \,  {\cal V}_{ \vec q}|0\ra \ , \ee
we expand the Neumann coefficients in Eq.~\eqref{n33tq} in a power series in $\vec q^{\, 2}$, retaining however
in each term the $\vec q^{\, 2}$-dependent powers of the energy in Eq.~\eqref{expq}. We then expand the exponential
rewriting the operator in Eq.~\eqref{vtq} as follows
\be  {\cal V}_{ \vec q} ={\cal W} \si [ 1 + \vec q^{\, 2} O_2 + \vec q^{\, 2} q^{i}  O^{i}_3 + ... + q^{i_1} ... q^{i_k} O^{i_1 ... i_k}_k + \dots
\de ] \ , \label{qseries} \ee
where the operator ${\cal W}$ is given by 
\be {\cal W} = \b \, {\cal P}_n  \, e^{\frac{1}{2} \sum_{k,l} W^{33}_{k l} a^i_{-k} a^i_{-l} + \sqrt{\frac{\ap}{2}}\sum_k 
W^3_k a^i_{-k} q^i}|0\ra \ , \ee
with
\be
W^{33}_{k l} =   \, 
N^{33}_{k l}(0) \,  e^{- \frac{\ap \vec q^{\, 2}}{4}\frac{k+l}{n} \log n} \ , \hspace{1cm} W^3_k = 
\frac{k^{\frac{k}{n}-1}}{\Gamma\si(1+\frac{k}{n}\de)} \, 
 e^{- \frac{\ap \vec q^{\, 2}}{4} \frac{k}{n} \log n}   \ . \ee
Here $N^{33}_{k l}(0)$ are the Neumann coefficients at zero transverse momentum
in~\eqref{v1} and  the  coefficients $W^{33}_{k l}$ coincide with them except for 
the Gaussian factor in $\vec q$. Note also that we kept in the exponential the
terms linear in $\vec q$ and in the open modes.

We can evaluate the contribution of the first term in the series following the same steps as at the beginning of this section.
We find
\ba  \widetilde{{\cal V}}_{ \vec b}|0\ra &\sim& \b \,\int \frac{d^{24-p}q}{(2\pi)^{24-p}}
\, e^{-i \vec b \vec q} \, {\cal P}_n \si [  e^{\frac{1}{2} \sum_{k,l} W^{33}_{k l} a^i_{-k} a^i_{-l} + \sqrt{\frac{\ap}{2}}\sum_k 
 W^3_k a^i_{-k}  q^i} \de]|0\ra \nb \\
&=& \b \,  \int \frac{d^{24-p}q}{(2\pi)^{24-p}}
\, e^{-i \vec b \vec q}  \, e^{- \frac{\ap \vec q^{\, 2}}{4}  \log n}  \, {\cal P}_n \si [
\, e^{\frac{1}{2} \sum_{k,l} N^{33}_{k l}(0) a^i_{-k} a^i_{-l} + \sqrt{\frac{\ap}{2}}\sum_k 
\, \frac{k^{\frac{k}{n}-1}}{\Gamma\si(1+\frac{k}{n}\de)}  \,   a^i_{-k} q^i}\de]|0\ra \nb \\
&=& \frac{1}{\si (\pi \ap \log \ap s \de)^{\frac{24-p}{2}}} \,   \, {\cal P}_n \si [
\, e^{-\frac{1}{\ap \log \ap s}\si(b^i + i  \sqrt{\frac{\ap}{2}}\sum_k 
 \frac{1}{k} a^i_{-k}  \de)^2}  \,   {\cal V}_{\vec 0} \de]|0\ra   \ . \ea
In the second passage we used that, due to the presence of the projector, the $\vec q^{\, 2}$-dependent powers of the energy
in the coefficients $W^{33}_{k_\a l_\a}$ and $W^{3}_{k_\a}$ combine in an overall Gaussian factor. In the last passage we  
approximated $k/n \sim 0$ in the operator multiplying $ {\cal V}_{\vec 0} $
since in impact parameter space the leading contributions due to this operator arise from modes
with $k \ll n$. 

The Fourier transform of the first term in the series in Eq.~\eqref{qseries} thus coincides with the vertex at zero transverse momentum up to
the insertion of an operator with an overall dependence on $b$ given by a Gaussian form factor 
\be e^{-\frac{b^2}{\ap \log \ap s}} \ . \label{lg} \ee
Note that the length scale in this form factor is  the string scale enhanced by a logarithm of the energy
\be \ap \log \ap s \ . \ee
This is due to the well-known logarithmic growth of a highly energetic string in the transverse directions. Precisely
as a consequence of this phenomenon, all the other contributions in the series in Eq.~\eqref{qseries} are subleading in the high-energy
limit and the term we have just evaluated gives the full answer. 

To verify this 
we do not need the precise form of the operator coefficients $O^{i_1 ... i_k}$,
all that is relevant is that each term is  multiplied
by the Gaussian factor in Eq.~\eqref{lg}. When we evaluate the Fourier transform of the other terms in the series, the powers
of the transverse momentum become derivatives with respect to the impact parameter and the resulting contribution
is then suppressed by inverse powers of $\log \ap s$. 
The final result is therefore
\be  \widetilde{{\cal V}}_{\vec b}|0\ra  = \frac{ 1}{\si (\pi \ap \log \ap s \de)^{\frac{24-p}{2}}} \,   {\cal P}_n \si[
\, e^{-\frac{1}{\ap \log \ap s}\si(b^i + i  \sqrt{\frac{\ap}{2}}\sum_k 
 \frac{1}{k} a^i_{-k}  \de)^2}    {\cal V}_{ \vec 0}\de] |0\ra  \ . \label{bvertex} \ee
This is the form of the vertex that we will use in actual calculations. It can also be written in a very suggestive
form if we interpret the Gaussian factor  as a squeezed state for the effective creation and
destruction operators
\be B^i = \frac{1}{\sqrt{\log n}} \sum_{k=1}^n \frac{a^i_k}{k} \ , \hspace{1cm}
 B^{i \dagger} =  \frac{1}{\sqrt{\log n}} \sum_{k=1}^n \frac{a^i_{-k}}{k} \ . \label{effB}\ee
In the high energy limit they satisfy 
\be [B^i, B^{j \dagger}] = \frac{\d_{ij}}{\log n}   \sum_{k=1}^n \frac{1}{k} \sim \d_{ij} \ . \ee
Since the squeezed states represent the position eigenstates in the oscillator basis of the Hilbert space
we can write
\be |b \ra_X = \frac{1}{\si(\sqrt{\pi}l_n\de)^{\frac{24-p}{2}}}\
e^{-\frac{b^2}{2l_n^2} - i \frac{\sqrt{2}}{l_n} B^\dagger b + \frac{1}{2} \si(B^\dagger\de)^2} \, |0\ra
= e^{- i b  \vec{P}} |0 \ra_X \ , \hspace{1cm} l^2_n = \ap \log n \ , \ee
where $|b\ra_X$ is an eigenstate of the position operators
\be 
X^i =  i \frac{l_n}{\sqrt{2}} \si(B^i - B^{i \dagger} \de ) = i \sqrt{\frac{\ap}{2}} \sum_{k=1}^n  \si(\frac{a^i_k}{k}  - \frac{a^i_{-k}}{k} \de ) 
\equiv  -  2 \sqrt{\ap}  \sum_{k=1}^n \frac{p_k^i}{k}\ , \label{xeff} 
\ee
with eigenvalue $b^i$ and the $ P^i$ are the corresponding momentum operators
\be P^i =  \frac{1}{\sqrt{2}l_n} \si(B^i + B^{i \dagger} \de ) =  \frac{1}{\sqrt{2\ap} \log \ap s}
 \sum_{k=1}^n  \si(\frac{a^i_k}{k}  + \frac{a^i_{-k}}{k} \de ) 
\equiv \frac{1}{ \sqrt{2} \ap \log \ap s} \, \sum_{k=1}^n  x_k^i \ . \label{peff} \ee
Here $x_k$ and $p_k$ are the position and momentum modes of the open string, as defined
in the two previous equations.
The state $|b\rangle_X$ is normalized in the standard way,
$\langle b | b' \rangle = \delta (b - b')$. We can then write 
\ba  \widetilde{{\cal V}}_{\vec b}|0\ra  &=& \frac{ e^{-\frac{b^2}{2 \ap \log \ap s}}}{\si (\pi \ap \log \ap s \de)^{\frac{24-p}{4}}} \,   {\cal P}_n 
   \,   {\cal V}_{ \vec 0} \,  |b \ra_X   \;. \label{vbd} \ea
This form of the closed-open vertex in impact parameter space is very natural. 
The appearance of the  squeezed state $|b\ra_X$ indicates that the closed-open transition
is proportional to a delta function $\d(X-b)$ for the effective position operator $X$. 
Its presence reflects the locality of the string interactions, the light-cone vertex
allowing only transitions between closed and open strings that overlap along their
entire length. This requires in particular that the
closed string should  touch the branes in at least one point   in order for the transition to take place.
The way that a closed string initially localized away from the location of the branes manages to touch
them and split is via quantum fluctuations in its position. 

From Eq.~\eqref{bvertex}  we see in fact that the transition happens with a non negligible amplitude as long as
$b^2 \le \ap \log \ap s$, that is for impact parameters that can be much larger than the maximal size $\ap M$ of a string of mass
$M$, which is kept fixed in the high energy limit. We can picture the transition as happening in two stages.
First the closed string becomes polarized, stretching towards the branes so that it touches them 
in at least one point, then it splits to form the final
open string. 
The Gaussian dependence on the impact parameter of the transition amplitude can then be related to the 
probability amplitude of finding the closed string in a configuration where it overlaps with the branes.
We will confirm this picture  in~\cite{classicalclop} where we will discuss the classical open string solutions that correspond
to the state created on the branes in the closed-open transition. 

Let us finally note that using  Eq.~\eqref{vbd} one can prove that states created at different values  of the impact parameter are orthogonal.
This follows from the fact that at high energy  ${\cal P}_n {\cal V}_0 |b\ra_X$ remains to a first approximation an eigenstate of the
transverse position opertors $X^i$. 

Using this vertex we can evaluate the discontinuity of the elastic amplitude in impact parameter space.
The calculation is most easily done representing the imaginary part as the product of the vertex at
impact parameter $b$ and the vertex at zero momentum, as in Eq.~\eqref{imgb}
\be  {\rm Im}{\cal A}_{TT}(s, b) =  \pi \ap \la 0 |{\cal V}^\dagger_{\vec 0} \widetilde{{\cal V}}_{\vec b}  | 0 \ra  \ . \ee
In order to evaluate
\be \la 0|  {\cal V}^\dagger_{ \vec 0}   \,  {\cal P}_n \si[
\, e^{-\frac{1}{\ap \log \ap s}\si(b^i + i  \sqrt{\frac{\ap}{2}}\sum_k 
 \frac{1}{k} a^i_{-k}  \de)^2}   \,   {\cal V}_{ \vec 0}\de] |0\ra 
\ , \ee
it is sufficient to notice that, compared with the calculation done for $t=0$, the additional terms in the previous formula
give contributions that 
are suppressed by powers of $\log \ap s$.  We then find
\be  {\rm Im}{\cal A}_{TT}(s, b) =  \pi \ap \b^2  \, \ap s\, \frac{e^{-\frac{b^2}{\ap \log \ap s}}}{\si (\pi \ap \log \ap s \de)^{\frac{24-p}{2}}}
\,  \ , \ee
reproducing Eq.~\eqref{imab}. Here the factor $\alpha' s \equiv n$ comes from the matrix element containing two vertices at $t=0$,
 as we have already seen in Eqs. (\ref{5.6}), (\ref{5.10}) and (\ref{conjsum}).

\subsection{Massless states}

We now discuss the absorption of a massless closed string state
\be |g_\e, \bar g_{\bar \e}\ra   = (\e  A_{-1})\, (\bar \e\bar A_{-1}) |0\ra \ , \ee
with polarization tensor $\e_{i}   \bar \e_{j}$.
The open state created by the absorption of a very energetic massless closed string
has an extremely simple description, simpler than the one we found for the absorption of a tachyon. 
When the transverse momentum vanishes it is given by
\be  {\cal V}_{\vec 0}  |g_\e, \bar g_{\bar \e}\ra  = 
 (\e a_{-1}) \, (a_{-n+1} D \bar \e) |0\ra \ . \label{mo3} \ee
In order to derive this result we have to deal with some technical complications, which arise because
our light-cone gauge is not very suitable to describe the right part of a closed massless
state with zero transverse momentum. The action of the reflection matrix $D^{\m\n}$ makes it equivalent to a state with $p^+ = 0$, see $\alpha_2$ in~\eqref{alpha1alpha2} with $N_c=1$ and $\vec p_t=0$. 
We could simply evaluate the transition amplitudes with $\vec p_t \ne 0$ and then take the limit $\vec p_t \rightarrow 0$. 
However it turns out that in order to have a well defined limit it is necessary to give a small mass $\m$ to the massless state and send
first $\vec p_t$ to zero and then $\m$ to zero. Otherwise the result of the limit would depend on the direction of $\vec p_t$,
as a consequence of the singular behaviour of
the Neumann coefficients $N^2_k$. A more detailed discussion of this subtlety can be found in 
Appendix \ref{checks}, where we make a direct comparison between a covariant amplitude and an amplitude
derived using the light-cone vertex and show that the two agree when the limit of zero transverse momentum is
taken as explained above. 

When acting on a massless state the general closed-open vertex becomes
\be V_{\vec p_t} |g_\e, \bar g_{\bar \e}\ra  = \b \, {\cal P}_n e^{Z_o+Z_{c,g}} |g_\e, \bar g_{\bar \e}\ra \ , \ee
where
\ba Z_{c,g} &=& N_{11}^{12}A_{1} D \bar A_{1} +N_{1u}^{23}  \bar A_{1} D a_{-u} +N^{13}_{1u} A_{1} a_{-u} + 
 N^1_1 \vec P A_1  +  N^2_1 \vec P D \bar A_1  \ , \nb \\
Z_o &=& \frac{1}{2} N^{33}_{kl} a_{-k} a_{-l}  +  N^3_u P a_{-u}\ . \ea
Therefore
\ba  V_{\vec p_t} |g_\e, \bar g_{\bar \e}\ra  & =&  \b \, {\cal P}_n \,  e^{Z_o}\si [ (\e D \bar \e)\,  N^{12}_{11} + 
  N^{23}_{1u} N^{13}_{1v} (\e a_{-v})\,  (a_{-u} D \bar \e)  +
 \, N^{23}_{1u} N^1_1  (a_{-u} D \bar \e)   \, (\e \vec P) \de . \nb \\
&+& \si .  \, N^{13}_{1u} N^2_1   (\e a_{-u}) \, (\bar \e D \vec P)
+  N^1_1 N^2_1   (\bar \e D \vec P)  (\e \vec P) \ \ \de]|0\ra \ .  \label{mo1} \ea
This expression simplifies considerably in the high-energy limit.
Let us set
\be \l = \frac{\ap}{4} \vec p_t^{\, 2} + \ap \m^2 \  , \ee
where $\m$ is the small mass introduced to regularize the  limit $\vec p_t \rightarrow 0$. 
For large $n$ the  Neumann coefficients in Eq.~\eqref{mo1} behave as follows
\be \label{eq:537}
N^{12}_{11}   \sim 1
 \ , \hspace{1cm}
N^{13}_{1 k } \sim  -
\frac{ 1}{ k - 1}\frac{\l}{n}   \,   \frac{k^{-\l \frac{k}{n}}}{\Gamma\si(1-\l \frac{k}{n}\de)}
 \ , \hspace{1cm} k \ne 1  \ , \hspace{1cm}
N^{13}_{1 1 } \sim  1  \ . \ee
Note the enhancement of the contractions between left modes and open modes when the mode
numbers coincide. We also have
\be
N^{23}_{1 k}   \sim 
 \frac{ n}{ n-\l k}  \,   \frac{k^{-\l \frac{k}{n}}}{\Gamma\si(1-\l \frac{k}{n}\de)}\ , \hspace{1cm}
N^{33}_{k l} \sim - \, \frac{ \l}{ n(k+l)} \,
 \frac{k^{-\l \frac{k}{n}}}{\Gamma\si(1-\l \frac{k}{n}\de)}  \frac{l^{-\l \frac{l}{n}}}{\Gamma\si(1-\l \frac{l}{n}\de)}  \ , \ee
and \be N^1_{1} \vec P \sim -  \sqrt{\frac{\ap}{2}}\vec p_t \ , \hspace{1cm}
 N^2_1 \vec P \sim  - \frac{n }{\l} \, \sqrt{\frac{\ap}{2}}\vec p_t\ , \hspace{1cm}
N^3_{k} \vec P \sim    \frac{1}{ k} \,   \,   \frac{k^{-\l \frac{k}{n}}}{\Gamma\si(1-\l \frac{k}{n}\de)}
 \sqrt{\frac{\ap}{2}}\vec p_t
\ . \label{n01} \ee
Focusing on the energy dependence of the Neumann coefficients, we can see that the fourth term  in Eq.~\eqref{mo1} would give a leading contribution of order $n$ which however disappears in the $\vec p_t \rightarrow 0$ limit. Similarly we can discard all other terms containing factors of $\vec P$ which vanish  in  the $\vec p_t \rightarrow 0$ limit thanks to the presence of the mass $\m$. Thus the next most relevant term is the second when $v=1$ and $u=n-1$, so as to use the last in~\eqref{eq:537}, and is of order $\sqrt{n}$ when written in terms of mode operators that commute to one. Therefore the leading terms at high energy are
\be {\cal  V}_{\vec p_t} |g_\e, \bar g_{\bar \e}\ra  \sim \b \, {\cal P}_n \, \left[ e^{Z_o}
 N^{13}_{11}   N^{23}_{1u} (\e a_{-1}) \, (a_{-u} D \bar \e)  \right]|0\ra \ .  \label{mo1he} \ee
Let us now take the limit $\vec p_t \rightarrow 0$ on $Z_o$.
Since $N^{33}_{kl}$ vanish when we
set $\m = 0$ we also have $e^{Z_o} \sim 1$. Taking into account the projector ${\cal P}_n$
and using that $N^{23}_{1,n-1} \sim n^{-\ap \m^2}$ we find the state in Eq.~\eqref{mo3}
\ba &&  {\cal V}_{\vec 0}  |g_\e, \bar g_{\bar \e}\ra  =  
 (\e  a_{-1}) \, (a_{-n+1} D \bar \e)  |0\ra \ . \label{mo3bis} \ea
This open state reproduces the imaginary part of the elastic amplitude at $t=0$ in the high-energy limit in Eq.~\eqref{imato}
\ba {\rm Im}{\cal A}_{gg} &=& \pi \ap  \la g_\z, \bar g_{\bar \z}|  {\cal V}_{\vec 0}^\dagger  {\cal V}_{\vec 0} |g, \bar g\ra 
=  \pi \ap \b^2 \, \e_s (D \bar \e)_r \z_{s'} (D \bar \z)_{r'} \la 0 |a_{n-1}^{r'} a_{1}^{s'} a_{-n+1}^r a_{-1}^s|0\ra \nb\\
&=&  \pi \ap \b^2 (\e \z) (\bar\e \bar\z) (n-1) \sim  \pi \ap \b^2  n  (\e \z) (\bar\e \bar\z)  \ , \ea
as one can see by using Eq. (\ref{beta}).

Let us now consider the leading terms in the expansion of $ {\cal V}_{\vec p_t}$ for small transverse momenta
and derive the absorption amplitude in impact parameter space. We shall focus on the leading term in Eq.~\eqref{imabo},
neglecting contributions suppressed by  additional powers of $\log \ap s$. The steps are essentially the same as for the
absorption of the tachyon. We first notice that as in that case the Neumann coefficients multiplying an open mode $a_{-k}$
have a dependence on the transverse momentum of the form $k^{-\frac{\ap}{4}\vec q^{\, 2} \frac{k}{n} \log n}$, that  
in impact parameter space gives again a Gaussian in $b$.  We can then neglect all the terms proportional
to $\vec P$ since they become derivatives $\p_b$ and correspond
to subleading terms\footnote{If we set $\m = 0$ before taking the limit of small transverse momentum
this would not be true for the  terms containing $N^2_1$ and up to two factors
of $\vec P$, since $N^2_1 \sim \frac{1}{\ap \vec p_t^{\, 2}}$. These terms would lead to a dependence of 
the open state on the direction along which  the transverse momentum
is sent to zero.} in an expansion in powers of $\frac{b^2}{\ap\log \ap s}$. 
Therefore at high energy and for small $\l$ the second (dominant) term in the right-hand-side of Eq.~\eqref{mo1} becomes
\ba   && {\cal  V}_{\vec p_t} |g_\e, \bar g_{\bar \e}\ra \sim - \b  \ \, {\cal P}_n \, \si [ e^{   N^3_k P_i  a^i_{-k}} \,
  N^{23}_{1 u} ( \e a_{-1}) \, (a_{-u} D \bar \e) \de   ] \, |0\ra   \ .  \label{mo2} \ea
Using that 
\be 
N^{23}_{1 u} \sim u^{-\l \frac{u}{n}} \ , \hspace{1cm} N^3_k \sim  \frac{k^{-\l \frac{k}{n}}}{k} \sqrt{\frac{\ap}{2}} \vec p_t
 \ , \ee
and following the same steps as in the discussion of the tachyon, we find in impact parameter space 
\ba  \widetilde{{\cal V}}_{\vec b} \, |g_\e, \bar g_{\bar \e}\ra  &=& \frac{1}{\si (\pi \ap \log \ap s \de)^{\frac{24-p}{2}}} 
\,   {\cal P}_n \si[\, e^{-\frac{1}{\ap \log \ap s}\si(b^i + i  \sqrt{\frac{\ap}{2}}\sum_k \frac{1}{k} a^i_{-k}  \de)^2}    \sum_{u=1}^\infty (\e a_{-1}) \, ( a_{-u}D\bar\e)  \de] |0\ra  \nb \\
&=& \frac{ e^{-\frac{b^2}{2 \ap \log \ap s}}}{\si (\pi \ap \log \ap s \de)^{\frac{24-p}{4}}} \,  {\cal P}_n 
   \,    \sum_{u=1}^\infty (\e a_{-1}) \, (\bar\e D a_{-u}) \,|b\ra_X\ , \ea
and
\ba {\rm Im}{\cal A}_{gg} &=&    \pi \ap  \la g_\z, \bar g_{\bar \z}|
{\cal V}^\dagger_{\vec 0} \, \widetilde{{\cal V}}_{\vec b}   |g_\e, \bar g_{\bar \e}\ra =
 (\e \z) (\bar\e \bar\z) \, {\rm Im}{\cal A}_{TT}  \ , \ea
in agreement with Eq. (\ref{2.25}).

\subsection{First massive level}

We turn now to the absorption of a massive string state belonging to the first   
massive level $N_c=2$, reviewed in Section~\ref{scattsec}. We will 
discuss the absorption of a closed state of the form $(L \otimes \bar L )$ or $(H \otimes \bar H)$. 
We will see that in both cases  the enhancement
of the $N^{13}$ Neumann coefficients in Eq.~\eqref{enhance} leads to a simple form for the open state.
The analysis of the absorption of $(H \otimes \bar H )$ will also show that 
in the closed-open vertex at high energy while the left components of the polarization tensor always appear
contracted with open modes (terms proportional to $N^{13}$), the right components
can appear either contracted with open modes (terms proportional to $N^{23}$) or
among themselves  (terms proportional to $N^{22}$). 

It is therefore not immediate to see how the product of two absorption amplitudes can  reproduce 
the imaginary part of the elastic amplitude in Eqs.~\eqref{imato} since the latter contains only 
 one type of  contraction, where the left (right) components of the polarization tensor  of one state
are contracted with the left (right) components of the other. We will argue that the coefficients
of all the other types of contractions that could appear in the product indeed vanish.

The simplification of the form of the open state due to the enhancement
of the $N^{13}$ Neumann coefficients and the above-mentioned cancellation 
are the two main new features that appear in the evaluation of the absorption of a massive
closed string. The analysis of the first massive level will then make the discussion of the general
case relatively straightforward. 

Let us consider first the state
\be |L_\e, \bar L_{\bar \e}\ra = \frac{1}{2} (\e A_{-2}) \, (\bar \e \bar A_{-2}) |0\ra \ . \ee
At high energy the Neumann coefficients become
\ba
N^{33}_{k l} &\sim&  - \frac{\w}{n}
 \frac{ 1}{ k+ l} \, \frac{k^{-\w \frac{k}{n}}}{\Gamma\si(1-\w \frac{k}{n}\de)} \, \frac{l^{-\w \frac{l}{n}}}{\Gamma\si(1-\w \frac{l}{n}\de)} \ , \hspace{1cm} N^{23}_{2 k} \sim  \frac{n}{\w}
  \frac{ 2 }{ 2- \frac{\w}{n}k}  \frac{k^{-\frac{k}{n}}}{\Gamma\si(1- \frac{k}{n}\de)}
\ , \nb \\
N^{12}_{22}  &\sim&  -  \frac{n}{\w}  \ , \hspace{1cm}
N^{13}_{2 k} \sim \frac{ \w}{ n (k-2) }
  \frac{k^{- \w\frac{k}{n}}}{\Gamma\si(1- \w\frac{k}{n}\de)} 
 \ , \hspace{0.4cm} k \ne 2 \ ,  \hspace{1cm}
N^{13}_{2, 2} \sim -\frac{ 1}{2}  \ ,  \\
N^1_2 \vec P &=& \frac{1}{2} \sqrt{\frac{\ap}{2}} \vec p_t \ , \hspace{1cm}
N^2_2 \vec P = -  \frac{n^2}{\w^2} \sqrt{\frac{\ap}{2}} \vec p_t \ , \hspace{1cm}
N^3_k \vec P =  \frac{k^{-\w\frac{k}{n}}}{\Gamma\si(1- \w\frac{k}{n}\de)} \, \frac{1}{k} \sqrt{\frac{\ap}{2}} \vec p_t \ , \nb
\ea
where $\w = 1+ \frac{\ap}{4} \vec p{}^{\, 2}_t$. 
For this example we find, limiting ourselves to $\vec p_t = 0$, 
\be Z_{c,L} = N^{12}_{22}A_{2} D \bar A_{2} +N^{23}_{2u} \bar A_{2} D a_{-u} + N^{13}_{2u} A_{2} a_{-u}  \  ,  \ee
and
\ba && V_{\vec 0}|L_\e, \bar L_{\bar \e}\ra = \b {\cal P}_n e^{Z_o}\si [2 (\e  D\bar \e)  N^{12}_{22} + 
2  N^{13}_{2v} N^{23}_{2u} (\e a_{-v}) \, (a_{-u} D \bar \e)  \de]|0\ra \ . \label{vsv} \ea
To identify which of the terms in the square brackets in Eq.~\eqref{vsv} gives the leading contribution at high energy
we need to keep track of the powers of the energy associated with the Neumann coefficients $N^{33}_{kl}$.
The main observation 
is that since these coefficients  scale like
\be N^{33}_{kl} \sim n^{-\w\si(\frac{k}{n}+\frac{l}{n}\de)-2}  \ , \ee
the insertion of $Q$ couples of open modes gives a contribution to the discontinuity of the elastic amplitude that scales like
\be n^{-2Q-2\w} n^{2Q-1} = n^{-1-2\w} = n^{1-2N_c-\frac{\ap}{2} \vec p^2_t}\ , \ee
where the power $n^{2Q-1}$ comes from the sum over all possible states. 
In order for a term in the expansion of $e^{Z_{c,\psi}}$ to contribute with the leading power of the energy
 we then need a power of $n^{2N_c}$ from the square of the corresponding Neumann coefficients and the sum over the possible states. 
In the case at hand we need a factor of $n^4$. 
The first term in Eq.~\eqref{vsv} is therefore always subleading while in the second we need to set $v=2$ so that
\ba && {\cal V}_{\vec 0}|L_\e, \bar L_{\bar \e}\ra  = - \b \, N^{23}_{2u} \, {\cal P}_n \left[ e^{Z_o} (\e a_{-2}) \, (a_{-u} D \bar \e)  \right]   |0\ra \ . \label{aL} \ea
The form of the open state is very simple. The left part of the closed polarization tensor is contracted
with  open modes with the same mode number as the left closed modes, as a result
of the enhancement of the coefficient $N^{13}_{22}$, Eq.~\eqref{enhance}. The  right part of the
closed polarization tensor is contracted with  open modes with mode number $u$
of order $n$ (coefficients $N^{23}_{2u}$). 

As a check of Eq.~\eqref{aL} let us evaluate the product
\be \la L_\z, \bar L_{\bar \z}| 
 {\cal V}_{\vec 0}^\dagger \, {\cal V}_{\vec 0}|L_\e, \bar L_{\bar \e}\ra = \z_{s'} (D\bar \z)_{r'} \e_s (D\bar \e)_r \,
\b^2 \, \sum_{u, v} \,  N^{23}_{2u}N^{23}_{2v}
\, \la 0|  a_{2}^{s'} a_{v}^{r'}  e^{Z^\dagger_o} {\cal P}_n e^{Z_o}  a_{-2}^s a_{-u}^r  |0\ra 
 \ . \ee
In order to do this, we expand the exponentials and use  the fact that  the open modes with mode number of order
$n$ in $Z_o$ commute with the open modes with mode number of order one. The direct contraction between
the oscillators without any $N^{33}$ insertion is subleading since the sum over $u$ is then restricted
to $u = n-2$ and $N^{23}_{2,n-2} \sim \frac{4}{n}$. The next term requires the matrix element
\be   \la 0|  a_{v}^{r'}   \, a^i_{k'}a^i_{l'} 
 \, a^j_{-k}a^j_{-l}   a_{-u}^r  |0\ra \ , \ee
which gives two inequivalent contractions between the transverse indices.
We find
\be  {\rm Im}{\cal A}_{L L}  \sim \pi \ap \b^2 n (\e \z) (\bar \e \bar \z  ) \si [96 \cdot I_{L, 1}+ 8 \cdot I_{L, 2} \de ] \ , \ee
where the explicit expressions for the integrals 
$I_{L, i}$, $i=1, 2$, can be found in Appendix \ref{details} together with a more detailed description of the
calculation. Substituting the numerical values
we obtain the first approximation to the imaginary part of the disc
\be  {\rm Im}{\cal A}_{L L}  \sim \pi \ap \b^2 \, n (\e \z) (\bar \e \bar \z  ) \, 0.841 \ . \ee
If we include the second term in the series, we obtain
\be {\rm Im}{\cal A}_{L L}  \sim  \pi \ap \b^2 \, n (\e \z) (\bar \e \bar \z  ) \, 0.994 \ . \ee
The matrix elements and the integrals required to evaluate this second term are again collected
in Appendix \ref{details} (see Eqs. (\ref{B.8}), 
(\ref{B.9}) and (\ref{B.10})).

We turn now to the state
\be |H_\e, \bar H_{\bar \e}\ra = \frac{1}{2}\e^{i}_{(1)} \e^j_{(2)} \bar \e^{k}_{(1)} \bar \e^l_{(2)} 
A_{-1}^i A_{-1}^j  \bar A_{-1}^k \bar A_{-1}^l |0\ra \ , \ee
with $\hat\e^{ij} \equiv \e^{i}_{(1)} \e^j_{(2)} $ and $\hat{\bar\e}^{kl} \equiv  \bar \e_{(1)}^{k} \bar \e_{(2)}^l$ symmetric tensors. 
The  Neumann coefficients with two open modes are the same as before
while those with two closed modes are
\be N^{11}_{11} \sim  \frac{\w}{2n}   \ , \hspace{1cm}
N^{12}_{11}  \sim  1
 \ , \hspace{1cm} N^{22}_{11} \sim 
 \frac{  n^2}{ 2\w^2}   
\ . \ee
The mixed open-closed coefficients are
\ba 
N^{23}_{1 k} &\sim&    \frac{ 1}{1-\w \frac{k}{n}} \,  \frac{k^{- \w\frac{k}{n}}}{\Gamma\si(1- \w\frac{k}{n}\de)}
   \ , \hspace{1.38cm}
k \ne  n 
 \ , \hspace{1cm}
N^{23}_{1 n} \sim    \frac{1}{ n}   \ , \hspace{0.4cm}
k =  n   \ , \nb \\
N^{13}_{1 k} &\sim& - \frac{ \w}{ n (k-1)} \, 
  \frac{k^{- \w\frac{k}{n}}}{\Gamma\si(1- \w\frac{k}{n}\de)} 
 \ , \hspace{0.8cm} k \ne 1 \ , \hspace{1.05cm}
N^{13}_{1 1 } \sim 1 \ . \ea
Finally
\be
N^1_1 \vec P = - \sqrt{\frac{\ap}{2}} \vec p_t \ , \hspace{1cm}
N^2_1 \vec P = -  \frac{n}{\w} \sqrt{\frac{\ap}{2}} \vec p_t \ , \hspace{1cm}
N^3_k \vec P =  \frac{k^{-\w\frac{k}{n}}}{\Gamma\si(1- \w\frac{k}{n}\de)} \, \frac{1}{k} \sqrt{\frac{\ap}{2}} \vec p_t \ . \ee
Let us discuss explicitly only the vertex with $\vec p_t = 0$. 
We  need to retain the following closed modes
\be Z_{c,H} = \frac{1}{2} N^{11}_{11}A_{1}^r A_{1}^r +  \frac{1}{2} N^{22}_{11}\bar A_{1}^r \bar A_{1}^r  
+ N^{12}_{11}A_{1}^r D \bar A_{1}^r +N^{23}_{1u}  \bar A_{1}^r D a_{-u}^r +N^{13}_{1u} A_{1}^r a_{-v}^r \  , \ee
and  terms up to the forth order in the expansion of $e^{Z_{c,H}}$  can contribute to the absorption amplitude.
We find
\ba &&  V_{\vec 0} |H_\e, \bar H_{\bar \e}\ra  =  \b \, {\cal P}_n \, e^{Z_o}\si [  (N^{12}_{11})^2 \si(\e_{(1)} D \bar \e_{(1)}\de) 
\si( \e_{(2)} D \bar \e_{(2)}\de) 
+  \frac{1}{2}  N^{11}_{11} N^{22}_{11}  \si(\e_{(1)} \e_{(2)}\de) \si(\bar \e_{(1)} \bar \e_{(2)}\de) \de  . \nb \\ \nb 
&+&  \frac{1}{2}  N^{22}_{11}N^{13}_{1u} N^{13}_{1v} \,
 \e_{(1)}^s \, \e_{(2)}^{t}  \si(\bar \e_{(1)} \bar \e_{(2)}\de) a_{-u}^s  a_{-v}^t   +
\frac{1}{2}  N^{11}_{11}N^{23}_{1u} N^{23}_{1v} \,
 \si(\e_{(1)} \e_{(2)}\de) (D \bar\e_{(1)})^{s} \, (D\bar \e_{(2)})^{t}  a_{-u}^s  a_{-v}^t  
\\ \nb  &+&    2 N^{12}_{11}N^{23}_{1u} N^{13}_{1v} 
\si(\e_{(1)} D \bar \e_{(1)}\de)  \e^t_{(2)} \, (D\bar \e)^{s}_{(2)}  a_{-u}^s   a_{-v}^t 
 \\  &+&  \si.  \frac{1}{2}  N^{23}_{1u} N^{23}_{1v} N^{13}_{1w} N^{13}_{1z} 
 \, \e^{t}_{(1)} \, \e^y_{(2)} \, (D \bar \e)^{r}_{(1)} \, (D \bar \e)_{(2)}^{s} \,
a_{-u}^r  a_{-v}^s   a_{-w}^t  a_{-z}^y \de ] |0\ra \ .  \ea
Let us analyze this expression and show that the leading contribution at high energy is due to  the first term in the second line and to the 
last term.
Substituting the explicit values of the Neumann coefficients in the first two terms we find contributions of the form
\be e^{Z_o}\si [\si(\e_{(1)} D \bar \e_{(1)}\de) \si( \e_{(2)} D \bar \e_{(2)}\de) 
 + \frac{n}{4}  \si(\e_{(1)} \e_{(2)}\de) \si(\bar \e_{(1)} \bar \e_{(2)}\de) \de ] |0\ra \ , \label{s20o} \ee
which are clearly subleading.
The next three terms are
\ba  &e^{Z_o}& \si [  \frac{n^2}{4} N^{13}_{1u} N^{13}_{1v} \,
 \si(\bar \e_{(1)} \bar \e_{(2)}\de)  \e_{(1)}^{s} \e_{(2)}^t a_{-u}^s  a_{-v}^t +
\frac{1}{4n}  N^{23}_{1u} N^{23}_{1v} \,
  \si(\e_{(1)} \e_{(2)}\de) (D\bar\e_{(1)})^{s} (D\bar\e_{(2)})^{t}  a_{-u}^s  a_{-v}^t  \de . \nb \\&+&   \si . 
2 N^{23}_{1u} N^{13}_{1v} \si(\e_{(1)} D \bar \e_{(1)}\de)  \e^t_{(2)} (D\bar \e_{(2)})^{s} a_{-u}^s   a_{-v}^t \de ] |0\ra \ . 
\ea
For generic values of the indices they are  subleading but when both $N^{13}$ factors are enhanced
the first  term scales with the power of the energy required to be relevant in the high energy 
limit. Finally also the term with four open modes gives a leading contribution only when both the $N^{13}$ factors are enhanced.
The result of this analysis is that
\ba 
{\cal V}_{\vec 0}|H_\e, \bar H_{\bar \e}\ra &=& \b \, {\cal P}_n \, e^{Z_o}\si [  \frac{1}{2}  \sum_{u,v}  N^{23}_{1u} N^{23}_{1v}  
 \e_{(1)}^{s} \e_{(2)}^t (D \bar \e_{(1)})^{r} (D \bar \e_{(2)})^{l} \, a_{-1}^s  a_{-1}^t \, 
a_{-u}^r  a_{-v}^l  \de . \nb \\ &+& \si .   \frac{n^2}{4} \,
 \si(\bar \e_{(1)} \bar \e_{(2)}\de) \e_{(1)}^{s} \e_{(2)}^t a_{-1}^s  a_{-1}^t  \, \de ] |0\ra   \ . \label{B.68} \ea
The open state has again a simple form. As before, the left part of the closed polarization tensor is contracted
with  open modes that coincide with the left closed modes, as a result
of the enhancement of the coefficient $N^{13}_{11}$, Eq.~\eqref{enhance}. However the indices of the right part of the
closed polarization tensor can be contracted either with  open modes
of order $n$ (coefficients $N^{23}_{1u}$) or among themselves (coefficient $N^{22}_{11}$). The 
latter terms give additional contributions to the closed-open couplings of closed states with  non-vanishing
traces for their right polarization tensor. 

Let us now consider the imaginary part of the elastic amplitude.
We proceed as we did for the tachyon, expanding the exponentials $e^{Z_o}$ and evaluating the first few terms in the 
series to show that they already account for most of the discontinuity. 
We shall use again the fact that low frequency open modes commute with the high frequency modes
in $Z_o$. We find
\ba 
\label{5.69}
 {\rm Im}{\cal A}_{H H}  &\sim & \pi \ap \b^2 \, 
\si(\hat\e_{ij} \hat\z_{ij} \de ) 
\frac{1}{8}\la 0|\si [ n^2 \,
 \hat{\bar \z}_{ll}  + 2  N^{23}_{1u'} N^{23}_{1v'}  (D \hat{\bar{\z}} D)_{r'w'}
 a_{u'}^{r'}  a_{v'}^{w'}    \de ]  e^{Z_o^\dagger} \nb \\
&& {\cal P}_n e^{Z_o}
\si [ n^2 \,
 \hat{\bar \e}_{kk} + 2  N^{23}_{1u} N^{23}_{1v}  (D \hat{\bar{\e}} D)_{rw} 
a_{-u}^r  a_{-v}^w    \de ]  |0\ra \ . 
\ea 
The first non-vanishing term in the series is
\be   
\label{5.70}
(\hat\e_{ij} \hat\z_{ij})  (\hat{\bar \e}_{kl} \hat{\bar \z}_{kl}) \,   \sum_{u, v}  \d_{u+v, n} \si ( N^{23}_{1u} N^{23}_{1v}\de )^2 \, u v 
\sim  (\hat\e_{ij} \hat\z_{ij})  (\hat{\bar \e}_{kl} \hat{\bar \z}_{kl}) \, n \,  \int_0^1 dx \frac{x^{1-2x}(1-x)^{-1+2x}}{\Gamma^2(2-x)\Gamma^2(1+x)} \ . \ee
The numerical value of the integral is 
\be I_{H, 1} =   \int_0^1 dx \frac{x^{1-2x}(1-x)^{-1+2x}}{\Gamma^2(2-x)\Gamma^2(1+x)} \sim 0.831797 \ , \ee
 and we see that, as in the previous cases, the first term already
accounts for most of the discontinuity. However, as shown in Eq. (\ref{B.13}) and also discussed below, the other two  contributions in Eq. (\ref{5.69}) turn out to give contractions among the polarizations that are not present in the imaginary part of the amplitude. In order to see if they cancel, we have included higher order terms that come from the expansion of the exponentials in Eq. (\ref{5.69}).
The detailed derivation of them is given in  Appendix~\ref{details}. The result is 
\ba  {\rm Im}{\cal A}_{H H}  &\sim& \pi \ap \b^2 \, n \, \si( \hat\e_{ij} \hat\z_{ij} \de ) \si [ \si (I_{H,1} + 12  I_{H, 4}  + 2  I_{H, 5}  \de )
\si( \hat{\bar \e}_{kl} \hat{\bar \z}_{kl} \de ) \de .  \label{imhho} \\
&+& \si .  \si(  -  \frac{1}{2}  I_{H, 2} + \frac{3}{2}   I_{H, 3}   +   \frac{1}{2} I_{H, 6}  - 6  I_{H, 7}  -
 \frac{1}{2}  I_{H, 8}  + 9  I_{H, 9}  + \frac{3}{4}  I_{H, 10} \de)  \si( \hat{\bar \e}_{kk} \hat{\bar \z}_{ll}  \de )  \de] \ , \nb  \ea
where the explicit expressions for the integrals $I_{H, i}$ and their numerical values are collected in the Appendix.
The main new feature is that the imaginary part seems to be 
proportional to two different contractions of the polarization tensors
\be 
\label{B.73}
   (\hat\e_{ij} \hat\z_{ij})  (\hat{\bar \e}_{kl} \hat{\bar \z}_{kl}) 
 \ , \hspace{1cm}   (\hat\e_{ij} \hat\z_{ij})   \hat{\bar \e}_{kk} \hat{\bar \z}_{ll} \ . \ee
The first contraction is the only one we expect according to the discussion in Section~\ref{scattsec} and therefore
the coefficient of the second one should vanish. 
A cancellation is indeed possible
since  the coefficient of the contraction $\hat{\bar \e}_{kk} \hat{\bar \z}_{ll}$ 
receives contributions from terms  with opposite sign
according to whether they contain an even or an odd number of coefficients $N^{33}_{kl}$. 
We do not have a general proof that the cancellation actually occurs but the explicit evaluation
of the higher order terms in Eq.~\eqref{imhho} seems to indicate that this is the case
since the sum of all the $I_{H, i}$ in the second line
of Eq.~\eqref{imhho}  
gives a result which is one order of magnitude smaller than
the individual terms. Indeed, substituting the numerical values  of the integrals in Eq.~\eqref{imhho}  we find
\ba && {\rm Im}{\cal A}_{H H}  \sim 
\pi \ap \b^2 \, n \, \si( \hat\e_{ij} \hat\z_{ij} \de ) \si [ 0.993 \si(\hat{\bar \e}_{kl} \hat{\bar \z}_{kl} \de ) 
- 0.004 \si( \hat{\bar \e}_{kk} \hat{\bar \z}_{ll}  \de )  \de] \ . \ea
It would be interesting to have a general proof of this cancellation to all orders
in the series expansion. 

So far our discussion of the closed-open vertex for the states of the first massive level has been limited
to $\vec p_t = 0$. As we did for the tachyon and the massless sector, we can derive the vertex in impact parameter
space  from the small $\vec p_t$ behaviour of the vertex in momentum space. The steps are the same and the result is again that
the vertex in impact parameter space is given by the vertex at zero momentum multiplied
by the squeezed state in the effective modes~\eqref{effB}. It is easy to check that 
it reproduces the leading term in the imaginary part of the elastic amplitude, Eq.~\eqref{imabo}. 

\subsection{Generic massive states}

The pattern observed in our study of the first massive level generalizes to arbitrary mass levels,
with the only requirement that the closed state be ultrarelativistic, $M \ll E$. We will find that, thanks to the
properties of the Neumann coefficients discussed in Section~\ref{henc}, even in this more general case it is still
possible to give a simple, explicit and systematic description of the open state created in the $s$-channel.
Consider a closed state $|\psi\ra$ at level $N_c$ 
\be |\psi\ra = \prod_{k=1}^\infty \frac{1}{\sqrt{n_k! k^{n_k}}} \si ( \e_{(k)} A_{-k} \de)^{n_k}
 \prod_{ l = 1}^\infty \frac{1}{\sqrt{ \bar n_{l}!  l^{ \bar n_{l}}}} 
\si (\bar \e_{(l)}  \bar A_{- l} \de)^{ \bar n_{l}} 
|0\ra\ , \ee
created by a collection
of left and right modes, $A^{i_k}_{-k}$ and $ \bar A^{j_{l}}_{- l}$, with multiplicity $n_k$ and $\bar n_l$ such that 
\be \sum_{ \bar k=1}^\infty k n_k = \sum_{ \bar l=1}^\infty l \bar n_{l} = N_c \ , \ee
and characterized by a collection of left and right polarization vectors $\e_{(k_\a)}^{i_{\a}}$, $\bar\e_{(l_\b)}^{j_{\b}}$
with $\a = 1, ..., n_k$ and $\b = 1, ..., \bar n_{l}$.
Here we are using the compact notation
\be \si ( \e_{(k)} A_{-k} \de)^{n_k} \equiv \prod_{\a=1}^{n_k} \e_{(k_\a)}^{i_{\a}}A^{i_{\a}}_{-k} \ . \ee
The action of the closed-open vertex 
on this state gives
\be {\cal V}_{\vec 0} |\psi\ra = \b \, {\cal P}_n \, e^{Z_o+Z_{c,\psi}} |\psi\ra \ , \label{vcog2}\ee
where, as before, $Z_o$  contains only open string modes 
while $Z_{c,\psi}$ both open and closed string modes, the latter restricted to those in the set used 
to define the closed state. 

Consider first the vertex with $\vec p_t = 0$. 
The action of $e^{Z_{c,\psi}}$ on the state gives a polynomial in the open modes of the same order as the
monomial in the closed modes that defines the closed state.
From the high-energy scaling of the Neumann coefficients discussed in Section \ref{henc}, it 
follows  that the leading contributions to the absorption amplitude are obtained when the indices
of the open and closed modes in the coefficients $N^{13}_{k u}$  coincide. 
The right closed modes are then either coupled to open modes of order $n$ through the coefficients
$N^{23}_{l u}$ or contracted among themselves using the coefficients 
$N^{22}_{k l}$. Finally one acts on the resulting polynomial with the operator
$e^{Z_o}$. The first few terms in the expansion of the exponential already give an accurate
description of the open state. 

We can then write the following general formula for  the open state created on the brane world-volume
when  the closed state $|\psi\ra$ is absorbed by the  D$p$-brane system
\ba && {\cal V}_{\vec 0} |\psi\ra = \b \, {\cal P}_n \, e^{Z_o} \,   \prod_{k=1}^\infty \sqrt{\frac{ k^{n_k}}{n_k!}} 
\si (N^{13}_{k k}\de)^{n_{k}}  \si ( \e_{(k)} a_{-k} \de)^{n_k}  \si [ \prod_{l=1}^\infty 
\sqrt{\frac{ l^{\bar n_{l}}}{\bar n_{l}!}} \prod_{\b=1}^{\bar n_l} N^{23}_{l u_{\b}} D \bar \e_{(l_\b)}^{j_{\b}}a^{j_{\b}}_{- u_{\b}}
 \de .  \nb \\ &+&  \si .  
 \sum_{r = 1}^\infty  \frac{\bar n_r}{2}(\bar n_r-1) N^{22}_{r r}\,\si( \bar \e_{(r_1)} \bar \e_{(r_2)}\de)\,
\sqrt{\frac{ r^{\bar n_{r}}}{\bar n_{r}!}}\prod_{\r=3}^{\bar n_r} N^{23}_{r u_{\r}} D \bar \e_{(r_\r)}^{j_{\r}}a^{j_{\r}}_{- u_{\r}}\,
 \prod_{l=1 \atop l \ne r}^\infty 
\sqrt{\frac{ l^{\bar n_{l}}}{\bar n_{l}!}} \prod_{\b=1}^{\bar n_l} N^{23}_{l u_{\b}}D \bar \e_{(l_\b)}^{j_{\b}}a^{j_{\b}}_{- u_{\b}}\,
\de .  \nb \\
&+& \si .   \sum_{r, s =1 \atop r \ne s}^\infty  \bar n_r \bar n_s 
N^{22}_{r s}\,\si( \bar \e^{j_{r,1}} \bar \e^{j_{s,1}}\de)\,
\sqrt{\frac{ r^{\bar n_{r}}}{\bar n_{r}!}}\prod_{\r=2}^{\bar n_r} N^{23}_{r u_{\r}} D \bar \e_{(r_\r)}^{j_{\r}} a^{j_{\r}}_{- u_{\r}}\,
\sqrt{\frac{ s^{\bar n_{s}}}{\bar n_{s}!}}\prod_{\s=2}^{\bar n_s} N^{23}_{s u_{\s}} D \bar \e_{(r_\s)}^{j_{\s}} a^{j_{\s}}_{- u_{\s}}\,
\de . \nb \\
&& \si . 
\prod_{l=1 \atop l \ne r, s}^\infty 
\sqrt{\frac{ l^{\bar n_{l}}}{\bar n_{l}!}} \prod_{\b=1}^{\bar n_l} N^{23}_{l u_{\b}} D \bar \e_{(l_\b)}^{j_{\b}}a^{j_{\b}}_{- u_{\b}}\, 
 + 
 \dots  \de ] |0\ra \ , \label{gos}  \ea   
where the dots stand for all the other possible pairings of the right closed modes. Although notationally cumbersome, 
the previous formula neatly summarizes the representation of the massive open state in our basis of light-cone modes. 
There is a subset of open modes that are contracted with the left polarization vectors and
copy precisely the left part of the closed state, in that they carry exactly the same indices.
There are then open modes of order $n$ contracted with the right polarization vectors, plus all possible terms that can be obtained 
by contracting couples of right polarization indices among themselves. 
Finally there are insertions of traces of open modes $a^i_{-u}a^i_{-v}$ from the expansion of the operator $e^{Z_o}$, as
for the absorption of a closed tachyon. 

We emphasize again that when one evaluates the imaginary part of the disk using the closed-open couplings
given by Eq.~\eqref{gos} 
only the term proportional to the contraction
\be \prod_{(k,\a)} \si(\z_{(k_\a)} \e_{(k_\a)} \de)  \, \prod_{(l,\b)} \si(\bar \z_{l_\b} \bar \e_{l_\b} \de)   \ , \ee
between the polarizations of the initial and the final state should remain, while all the terms containing one or more factors
of the form 
\be \si(\bar \z_{(r_\r)} \bar \z_{(s_\s)}\de) \si( \bar \e_{(r'_{\r'})}\bar \e_{(s'_{\s'})}\de)   \ , \ee
should cancel. We do not have a general proof that this is the case. 

Finally, the vertex in impact parameter space is given by the vertex at zero momentum multiplied
by the squeezed state in the effective modes in Eq.~\eqref{effB}
and reproduces the leading term in the imaginary part of the elastic amplitude, Eq.~\eqref{imabo}.

\section{Conclusions}
\label{conclusions}

In this paper we have started the analysis of the absorption of a light, very energetic string by a stack of D$p$-branes and of the consequent excitation of the latter. The final aim of this study is to show how such a complicated process can be described in terms of a unitary $S$-matrix, thereby generalizing to the absorption regime what has been achieved so far in the scattering regime (including tidal excitation of the closed string).

This problem, in general, is  a very complicated one. It becomes tractable, however, by appropriately restricting the kinematic regime under scrutiny as explained in detail in the introduction. This allows, on the one hand, to ignore closed-string loops (and therefore gravitational bremsstrahlung) and, on the other hand, to neglect higher corrections to the leading eikonal. Under these approximations one expects to be able to construct a unitary $S$-matrix in a Hilbert space consisting of a single (but possibly excited) closed string and of an arbitrary number of open strings living on the brane world-volume.

Here we have considered the first term of this eikonal resummation, namely the tree-level (disk) approximation, postponing to further work
 \cite{clopunitary} the full unitarization program. In this approximation also the open-string Hilbert space  contains a single (in general highly excited) string.
A very encouraging outcome of our analysis has been the emergence of a simple description of the relevant states in the open sector. This makes one hope that resummation of the eikonal series will not be that hard. At the same time, the simple properties of the open sector can be given a classical (or semi-classical) interpretation \cite{classicalclop}, suggesting that a semiclassical treatment of the whole series could be sufficient for yielding a unitary $S$-matrix.

We conclude by mentioning that the price to pay for making the problem tractable is to work at sufficiently small string coupling for the characteristic radius of the brane-induced geometry $R_p$ to be smaller than the string length $l_s \sim \sqrt{\alpha'}$ enhanced by a logarithmic factor.
In that limit we expect the dominantly produced open strings to be heavy. Making contact  with the AdS/CFT correspondence 
\cite{Maldacena:1997re}, \cite{Witten:1998qj} in the supergravity approximation will unfortunately require the opposite limit of a large $R_p$ in string units. Such a regime appears still far from what our present computational technology can achieve.

\vspace{7mm}
\noindent {\large \textbf{Acknowledgements} }\\

G.D. thanks the Galileo Galilei Institute for Theoretical Physics for hospitality and the INFN for partial support 
while this research was being carried out.  G.D. also gratefully
acknowledges the hospitality of the Department of Applied Mathematics
and Theoretical Physics (Cambridge) at various times during the completion of this work.
P. D. thanks  C. Maccaferri for discussions on the open-closed string vertex. 
The research of R. R. is partially supported by STFC (Grant ST/L000415/1, {\it String theory, gauge theory \& duality}).
G.V. would like to acknowledge the hospitality of the  
Kavli Institute for Theoretical Physics, University of California, Santa Barbara (research supported in part by the 
National Science Foundation under Grant No. NSF PHY11-25915)  
where part of this research has been carried out.

\begin{appendix}

\section{The light-cone vertex and the covariant amplitudes}
\label{checks}

In this Appendix we make a direct comparison between the closed-open transition amplitudes evaluated
using the light-cone vertex and those derived using the covariant formalism. 
The simplest closed-open transition amplitudes that can be evaluated using the covariant formalism are 
those to open states belonging to the leading Regge trajectory whose vertex operators are
\be {\cal V}_{S_n} = \frac{g_o}{\sqrt{n!}} \si [ \frac{-i}{\sqrt{2\ap}} \de ]^n
\prod_{i=1}^n \z_{\a_i} \, \p X^{\a_i} e^{i p_o X} \ , \hspace{1cm} \ap m^2 = n - 1 \ , \ee
where we follow the notation introduced in Sections~\ref{scattsec} and~\ref{abs}. 
Since transitions to the tachyon and the vector are forbidden by the kinematics we only need $n \ge 2$. 

As in Section~\ref{abs}, when we evaluate an amplitude using the covariant formalism we write
\be p_c = (E, \vec 0_p, \vec p_c) \ , \ee
where $\vec p_c$ belongs to the $(25-p)$-dimensional space transverse to the
branes. We also define the tensor $\d_\perp$ as the Kronecker delta in the spatial directions orthogonal
to $\hat p_c$, where $\hat p_c$ is the unit vector $\hat p_c = \frac{\vec p_c}{|\vec p_c|}$.

When we evaluate an amplitude using the closed-open vertex 
 we write
\be p_c = (E, \vec 0_p, \vec p_t, p) \ . \ee
Here $\vec p_t$ belongs to  the $(24-p)$-dimensional space transverse to the
branes and to the direction $x^{25} \equiv z$.

\subsection{Tachyon to leading Regge}

When the initial closed state is a tachyon we find
\be B_{T, S_n}(\hat p_c) = \b \frac{(-1)^n}{\sqrt{n!}}  \si (  \frac{n+3}{2} \de)^{\frac{n}{2}} \prod_{i=1}^n \si( \z_{i} \hat p_c\de ) \ , 
\label{btlr}\ee
where we used that $\ap \vec p_c{}^2 = n +3$. 
Only the polarization with all  indices longitudinal is excited. 
For instance for $n=2$ the polarization tensor of the state that is excited is
\be \z_{{\rm lon}, 2} = \frac{1}{5\sqrt{24}} \si ( - \d_\perp + 24 \hat p_c \hat p_c \de ) \ ,  
\label{pol2} \ee
and for $n=3$ 
\be \z_{{\rm lon}, 3} = \frac{1}{9\sqrt{8}} \si ( \d_\perp \hat p_c + \d_\perp \hat p_c + \d_\perp \hat p_c - 24 \hat p_c \hat p_c \hat p_c \de ) \ . \ee
At level $n =2$ and $n=3$ the states of the leading Regge trajectory give all the physical  states and
therefore we can use the amplitudes in Eq.~\eqref{btlr} to reconstruct the  discontinuity of the elastic amplitude
at the corresponding energies. Consider the case $n=2$. 
To evaluate the imaginary part of the elastic amplitude we write
\ba {\rm Im} A &=& \pi \ap \sum_{\z} B_{T, S_2}(\hat p_1) B_{T, S_2}(\hat p_2) 
= \pi \ap \b^2 \frac{25}{8} \sum_{\z} \si( \z_{ij} \hat p^i_1 \hat p^j_1 \de)
 \si( \z_{kl} \hat p^k_2 \hat p^l_2 \de)  \\
&=& \pi \ap \b^2 \frac{25}{8}\si [ \si(\hat p_1 \hat p_2\de)^2 - \frac{1}{25}\de ]
= \frac{  \pi \ap \b^2}{8}\si [ 25 \cos^2 \th - 1\de ] = \pi \ap \b^2 \si ( 3 + \frac{5}{2} y + \frac{y^2}{2}\de )\nb
\ , \ea
where $\th$ is the angle between $\hat p_1$ and $\hat p_2$. The result agrees with Eq.~\eqref{eq:polesn}.
We used that the sum over a complete set of traceless symmetric tensors of rank two in $25$
spatial dimensions gives
\be \sum_\z \z_{ij} \z_{kl} = \frac{1}{2}\si ( \d_{ik} \d_{jl} + \d_{il} \d_{jk} \de ) - \frac{1}{25} \d_{ij} \d_{kl} \ , \ee
and that
\be \sin^2 \frac{\th}{2} = - \frac{y}{\ap(E^2-M^2)} \ , \hspace{1cm} y = \frac{\ap}{4} t \ . \ee
The cases $n \ge 3$ can be treated in a similar way. 

Let us now study the transition from the tachyon to the first massive level using the light-cone vertex. 
The closed-open vertex gives in this case
\be V_{\vec p_t}|0\ra = \b \si[\frac{1}{2} N^{33}_{11} a^i_{-1}a^i_{-1} + \sqrt{\frac{\ap}{2}}N^3_2 \a_3 p^i_t \, a^i_{-2} 
+ \frac{\ap}{4} \si( N^3_1 \a_3 \de)^2 p^i_t p^j_t \, a^i_{-1}  a^j_{-1}  \de]|0\ra \ ,\ee
where
\be N^{33}_{11} = - \frac{1-\r^2}{8}   \ , \hspace{1cm} N^3_2 \a_3 = \frac{\r}{2} 
 \ , \hspace{1cm} N^3_1 \a_3 =  1 \ , \hspace{1cm}  \r \equiv \frac{p}{E} = \sqrt{5-\ap \vec p_t^2}\ . \ee
More explicitly
\be V_{\vec p_t}|0\ra = \b \si[\frac{4-\ap \vec p_t^2}{16} a^i_{-1}a^i_{-1} + 
\frac{ \sqrt{5-\ap \vec p_t^2}}{2} \, \sqrt{\frac{\ap}{2}}p^i_t \, a^i_{-2} 
+ \frac{\ap}{4}  p^i_t p^j_t \, a^i_{-1}  a^j_{-1}  \de]|0\ra \ . \ee
Therefore the covariant state with the polarization tensor in Eq.~\eqref{pol2} corresponds to the
normalized state
\be  | \z_{{\rm lon}, 2}, \vec p_t\ra = \frac{1}{\sqrt{3}}\si[\frac{4-\ap \vec p_t^2}{16} a^i_{-1}a^i_{-1} + 
\frac{ \sqrt{5-\ap \vec p_t^2}}{2} \, \sqrt{\frac{\ap}{2}}p^i_t \, a^i_{-2} 
+ \frac{\ap}{4}  p^i_t p^j_t \, a^i_{-1}  a^j_{-1}  \de]|0\ra \ , \label{z2r} \ee
which for $\vec p_t = 0$ (i.e. when  the light-cone gauge is aligned to $\hat p_c$) reduces to
\be  | \z_{{\rm lon}, 2}, 0\ra = \frac{1}{\sqrt{48}} \, a^i_{-1}a^i_{-1} |0\ra \ . \label{z2} \ee
It is amusing to verify that the state in Eq.~\eqref{z2r} can be obtained from the state in Eq.~\eqref{z2}
by a rotation of  an angle $\f$ with $\sin \f = \frac{|\vec p_t|}{|\vec p_c|}$ in the plane 
$(\hat p_c, \hat p_t)$, as it should. If we call the rotation plane the plane $(z, y)$, the rotation operator is
\be R_{zy}(\f) = e^{-i \f J^{zy}} \ , \hspace{1cm} J^{zy} =  \frac{i}{2\sqrt{\ap}|p_o^+|} 
\sum_{k=1}^\infty \frac{1}{k} \si ( a^y_{-k} L_k - L_{-k} a^y_k \de ) \ , \ee
where $L_k = \frac{1}{2} \sum_{l \ne 0} a^i_{k-l} a^i_l$.
Let us consider for simplicity an infinitesimal rotation. When $\f \ll 1$ we find from Eq.~\eqref{z2r} 
\be  | \z_{{\rm lon}, 2}, \vec p_t\ra \sim \frac{1}{4\sqrt{3}}\si[ a^i_{-1}a^i_{-1} + 5 \sqrt{2} \f 
\, a^y_{-2}   \de]|0\ra \ .  \ee
The term linear in $\f$ coincides with the effect of an infinitesimal rotation on the state with $\vec p_t = 0$
\ba &&-i \f J^{zy}  | \z_{{\rm lon}, 2}, 0\ra  = \frac{1}{4\sqrt{3}} \frac{\f}{\sqrt{2}}
\sum_{k=1}^\infty \frac{1}{k} \si [ a^y_{-k} L_k - L_{-k} a^y_k,  a^i_{-1}a^i_{-1}\de ] |0\ra \\
&=&\frac{1}{4\sqrt{3}} \frac{\f}{\sqrt{2}}  \si ( \frac{1}{2} \si [ a^y_{-2} L_2,  a^i_{-1}a^i_{-1}\de ] -
2 L_{-1}a_{-1}^y \de )|0\ra =  \frac{1}{4\sqrt{3}} \frac{\f}{\sqrt{2}}  \si (12-2 \de ) a_{-2}^y|0\ra\ . \nb \ea
We can finally reconstruct the imaginary part of the elastic amplitude for $\ap E^2 = 1$. 
Since we are not in the high energy limit we work in the brick wall frame
defined in Eq.~\eqref{popenpclosed}. Using Eq.~\eqref{img1} we find 
\ba  {\rm Im} A &=& \pi \ap \b^2 \si[ 12 \si(N^{33}_{11}\de)^2
+ \frac{y^2}{8} \si( N^3_1 \a_3 \de)^4 - \frac{y}{2}  N^{33}_{11}  \si( N^3_1 \a_3 \de)^2 
-  y \si( N^3_2 \a_3 \de)^2\de] \nb \\ &=& \pi \ap \b^2 \si( 3 + \frac{5}{2} y + \frac{y^2}{2}\de )\ , \ea
in agreement with Eq.~\eqref{eq:polesn}.

\subsection{Massless to leading Regge}

When the initial closed state is a massless state with vertex operator
\be {\cal V}_g = - \frac{\kappa}{2\pi} \frac{2}{\ap} \, \e_\m \bar \e_\n \, 
 \p X^{\m}  \bar \p X^\n \,  e^{i p_c X} \ , \hspace{1cm} \ap M^2 = 0 \ , \ee
we find
\ba && B_{g, S_n} = - \b  \frac{1}{\sqrt{n!}}   \frac{2}{\ap} \si [ \frac{-i}{\sqrt{2\ap}} \de ]^n 
|z-\bar z|^{2-n}|z-x|^{2n} \, \e_\m  \bar \e_\n \si \{- \frac{\ap}{2} D^{\m\n}  \frac{1}{(z-\bar z)^2} L_n \de .  \\
&-& i \ap \frac{D^{\n \a_1}\z_{\a_1}}{(\bar z-x)^2} \si (\frac{\ap}{2} \frac{Dp_c^\m}{z-\bar z} + \ap \frac{p_o^\m}{z-x} \de ) n L_{n-1}
+ i \ap \frac{\h^{\m \a_1}\z_{\a_1}}{(z-x)^2}  \si (\frac{\ap}{2} \frac{Dp_c^\n}{z-\bar z} + \ap \frac{p_o^\n}{x-\bar z} \de ) n L_{n-1} \nb \\
&+&\ap^2 \frac{D^{\n \a_1}\z_{\a_1}\h^{\m \a_2}\z_{\a_2}}{(z-x)^2(\bar z - x)^2} n(n-1) L_{n-2}
+ \si .  \si (\frac{\ap}{2} \frac{Dp_c^\m}{z-\bar z} + \ap \frac{p_o^\m}{z-x} \de ) 
\si (\frac{\ap}{2} \frac{Dp_c^\n}{z-\bar z} + \ap \frac{p_o^\n}{x-\bar z} \de ) L_n \de \}\ , \nb \ea
where 
\be L_k = \si( \frac{z-\bar z}{|x-z|^2}  \de)^{k} \, \prod_{i=n-k+1}^{n}\si (  - i \ap |\vec p_c| \, \z_i \hat p_c \de) \ . \ee
As a basis for the polarizations of the massless closed string it is natural to use
 $24$ spacelike vectors orthogonal to the time direction $ \hat t$ and to $\hat p_c$. The same $24$ spacelike vectors
together with $\hat p_c$ provide a basis for the polarizations of the massive open state.
In this basis 
\be \e Dp_c =  \bar \e Dp_c = \e p_o = \bar \e p_o = 0 \ , \ee
and the previous expression reduces to
\be B_{g, S_n} 
=  \b  \frac{(-1)^{n+1}}{\sqrt{n!}} \, \si( \frac{n-1}{2} \de)^{\frac{n}{2}}
 \, \e_\m  \bar \e_\n \si [ D^{\m\n} \z_1 \hat p_c \z_2 \hat p_c  
+ 2  \, n \, \h^{\m \a_1}\z_{\a_1} D^{\n \a_2}\z_{\a_2}
 \de ] \, \prod_{i = 3}^{n} \si( \z_{i} \hat p_c\de )   \ , \ee
where we used that $\ap  \vec p_c {}^2 = n - 1$. Let us analyze in detail the
transition to the first massive level $(n=2)$
\ba B_{g, S_2} 
&=& - \b  \frac{1}{2\sqrt{2}} \,
 \, \e_\m  \bar \e_\n \si [ D^{\m\n} \z_{\r\s} \hat p_c^\r  \hat p_c^\s  
+ 4 \, \z^{\m}{}_{\a}D^{\a\n} \de ] \,    \ , \ea
where $\z_{\r\s}$ is symmetric and traceless.
We discuss separately the excitation of an open string with polarization transverse
to $\hat p_c$ and parallel to $\hat p_c$
\ba B_{g, S_2, {\rm tr}} 
&=& -   \sqrt{2} \b  \, \e_\m  D\bar \e_\n  \z^{\m\n} \ , \hspace{1cm}    B_{g, S_2, {\rm lon}} 
= -   \frac{\b}{\sqrt{12}}  \, \e_\m  \bar \e_\n  D^{\m\n}   \ , \ea
and analyze in turn the absorption of a dilaton $\f$, a graviton $G$ and a Kalb-Ramond field $B$. 
The dilaton can only excite longitudinally polarized states with amplitude~\footnote{Here $p$ is the dimension
of the D$p$-branes.}
\be B^\f_{g, S_2, {\rm lon}} =     \frac{\b}{\sqrt{72}} (12-p) \ . \ee
The graviton and the Kalb-Ramond field can only excite states with transverse polarization. 
The graviton can be absorbed when its indices are both parallel  to the brane or both
orthogonal to the brane and the collision axis
\be  B^{G_{\parallel, \parallel}}_{g, S_2,{\rm tr}} =  -   \sqrt{2} \b \ , \hspace{1cm} 
B^{G_{\parallel, \perp}}_{g, S_2,{\rm tr}} =     0
 \ , \hspace{1cm} B^{G_{\perp, \perp}}_{g, S_2,{\rm tr}} =     \sqrt{2} \b \ , \ee
while the Kalb-Ramond field  can be absorbed when one of its indices is parallel  to the brane and the other is
orthogonal to the brane and the collision axis
\be  B^{B_{\parallel, \parallel}}_{g, S_2,{\rm tr}}  =  0 \ , \hspace{1cm} B^{B_{\parallel, \perp}}_{g, S_2,{\rm tr}} =     -   \sqrt{2} \b
 \ , \hspace{1cm} B^{B_{\perp, \perp}}_{g, S_2,{\rm tr}} =    0 \ . \ee
If the massive polarization is longitudinal we find
\ba B_{g, S_2, l} 
&=& -   \frac{\b}{\sqrt{12}}  \, \e_\m  \bar \e_\n  D^{\m\n}    \ . \ea
We see that only the dilaton couples since all the other closed polarizations are traceless~\footnote{Here $p$ is the dimension
of the D$p$-branes.}
\be B_{g, S_2, l} =   \frac{\b}{\sqrt{12}} \frac{24-2p}{\sqrt{24}} =   \frac{\b}{\sqrt{72}} (12-p) \ . \ee
The basis of physical polarizations that we used in the previous discussion coincides with the basis of
light-cone polarizations for a massless state with $\vec p_t = 0$.
In this case the vertex gives
\ba && V_{0} |g_\e, \bar g_{\bar \e}\ra = \b \si[ N^{13}_{11} N^{23}_{11} \e_i D\bar \e_j a_{-1}^i a_{-1}^j
+ \frac{1}{2} \e_k D\bar \e_k N^{12}_{11} N^{33}_{11} a_{-1}^i a_{-1}^i \de ]\ , \ea
where 
\ba N^{13}_{11} &=& - \frac{1+\r}{2} = - 1 \ , \hspace{1cm} N^{23}_{11} = N^{12}_{11}  = 1 \ , \hspace{1cm}
N^{33}_{11} =  - \frac{1-\r^2}{8} = 0\ , \ea
since $\r = \frac{p}{E} = 1$. Therefore
\ba && V_{0} |g_\e, \bar g_{\bar \e}\ra = - \b  \e_i D\bar \e_j a_{-1}^i a_{-1}^j \ . \label{aavog} \ea
Consider now the following basis of light-cone states for the first massive level
\be |\chi_{\rm tr} \ra = \frac{\w_{ij}}{\sqrt{2}} \, a_{-1}^i \, a_{-1}^j \ ,\hspace{1cm} 
 |\chi_{\rm lon} \ra = \frac{1}{\sqrt{48}} \, a_{-1}^i \, a_{-1}^i \ ,\ee
with $\w_{ij}$ a symmetric traceless tensor in $24$ space directions. 
Then we find
\ba \la \chi_{\rm tr} | V_{0} |g_\e, \bar g_{\bar \e}\ra &=& - \sqrt{2} \b \e_i D \bar \e_j \w^{ij} \ , \nb \\
 \la \chi_{\rm lon} | V_{0} |g_\e, \bar g_{\bar \e}\ra &=& - \frac{\b}{\sqrt{12}}\e_i  \bar \e_j D^{ij} \ , \ea
in perfect agreement with the covariant amplitude.

Let us now consider $\vec p_t \ne 0$. In this case the basis of light-cone polarizations is different from the 
basis naturally associated to the states in the covariant calculation. The vertex gives
\ba && V_{\vec p_t} |g_\e, \bar g_{\bar \e}\ra = \b \si[ N^{13}_{11} N^{23}_{11} \e_i D\bar \e_j a_{-1}^i a_{-1}^j
+ \a^2_3 N^1_1 N^{23}_{11} N^3_1  \frac{\ap}{2} \si(\e \vec p_t\de) D\bar \e^i p^j_t a^i_{-1} a^j_{-1} \de . \nb \\
&+&  \a_3 N^1_1 N^{23}_{12} \sqrt{\frac{\ap}{2}} \si(\e \vec p_t\de) D\bar \e^i a^i_{-2}+
 \a^2_3 N^2_1 N^{13}_{11} N^3_1 \, \frac{\ap}{2} \si(D\bar \e \vec p_t\de)  \e^i p^j_t a^i_{-1} a^j_{-1} \nb \\
&+& \a_3 N^2_1 N^{13}_{12}\, \sqrt{\frac{\ap}{2}} \si(D\bar \e \vec p_t\de)  \e^i  a^i_{-2} 
+ \frac{1}{2}\si (\e_k D\bar \e_k N^{12}_{11} + \a^2_3 N^1_1 N^2_1  \frac{\ap}{2} \si(\e \vec p_t\de)  \si(D\bar \e \vec p_t\de)  \de ) \nb \\
&& \si . \si( N^{33}_{11} a_{-1}^i a_{-1}^i + \frac{\ap}{2} \a^2_3 \si ( N^3_1 \de)^2 p^i_t p^j_t a^i_{-1} a^j_{-1}
+ \sqrt{2\ap} \a_3 N^3_2 p^i_t a^i_{-2} \de ) \de ]\ , \ea
where 
\ba N^{13}_{11} &=& - \frac{1+\r}{2} \ , \hspace{1cm} N^{23}_{11} = N^{12}_{11}  = 1 \ , \hspace{1cm}
N^{33}_{11} =   - \frac{1-\r^2}{8} \ , \nb \\
\a_3 N^1_1 &=& - \frac{2}{1+\r}  \ , \hspace{1cm}
\a_3 N^2_1 = - \frac{2}{1-\r}   \ , \hspace{1cm}
\a_3 N^3_1 = 1  \ , \nb \\
N^{13}_{12} &=& - \frac{1-\r}{2}  \ , \hspace{1cm} N^{23}_{12} = \frac{1+\r}{2} \ , \hspace{1cm}
\a_3 N^{3}_{2} =   \frac{\r}{2} \ , \ea
and 
\be \r = \frac{p}{E} = \sqrt{1-\ap \vec p_t {}^{2}} \ . \ee
In the limit $\vec p_t \rightarrow 0$ we find
\be V_{\vec p_t} |g_\e, \bar g_{\bar \e}\ra \rightarrow  - \b  \e_i \si ( D\bar \e_j -  2 \si(D\bar \e \hat p_t\de)  \hat p^j_t  \de ) a_{-1}^i a_{-1}^j \ . \ee
We see that the result of the limit depends on the direction of $\vec p_t$. It does not agree with Eq.~\eqref{aavog} 
and does not reproduce the covariant amplitudes. 
The correct limit is found if we give a small mass $\m$ to the closed state setting
\be \r(\m) = \frac{p}{E} = \sqrt{1-\ap \si(\vec p_t{}^{2}+\m^2\de)} \ , \ee
and then send $\vec p_t$ to zero before removing the mass.

\section{On the calculation of ${\rm Im}{\cal A}$ from the vertex}
\label{details}

In this Appendix we describe in more detail how to derive the imaginary part of the disk 
starting from the closed-open vertex. We will discuss the absorption of the massive states
$|L,\bar L\ra$ and $|H,\bar H\ra$, already analyzed in Section~\ref{heco}.
It is convenient to define
\be \s_\w(k) \equiv \frac{k^{ - \w \frac{k}{n}}}{\Gamma \si( 1 - \w \frac{k}{n}\de)} \ , 
\hspace{1cm} \s_\w(x) \equiv \frac{x^{-\w x}}{\Gamma \si( 1 - \w x\de)} \ , 
\hspace{1cm}  \w = N - 1 + \frac{\ap}{4} \vec p_t^{\, 2}\ . \ee
We also introduce the following compact notation for the integration over the $n$-simplex  
\be \int d\m_n  \equiv \int_0^1 dx_1 ... \int_0^1 dx_n \d\si(\sum_{i=1}^n x_i - 1 \de ) \ . \ee
Let us begin with the evaluation of the first few terms in the series of the imaginary
part of the elastic amplitude for the state $|L,\bar L\ra$. 
The first contribution is 
\ba &&  {\rm Im}{\cal A}_{L L}  \sim
 \frac{\pi}{4} \ap \b^2 \,  \z_{s'} D\bar \z_{r'} \e_s D\bar \e_r \sum_{u, k, l \atop
v, k', l'} \d_{u+k+l, n} N^{23}_{2u} N^{23}_{2v}N^{33}_{kl}N^{33}_{k'l'}
 \la 0| a_2^{s'} a_{v}^{r'}   \, a^i_{k'}a^i_{l'} 
 \, a^j_{-k}a^j_{-l}   a_{-u}^r  a_{-2}^s|0\ra \nb \\
&=& \frac{\pi}{2} \ap \b^2 \,  \si ( \z \e \de )  D\bar \z_{r'}  D\bar \e_r   \sum_{u, k, l \atop
v, k', l'} \d_{u+k+l, n}N^{23}_{2u} N^{23}_{2v}N^{33}_{kl}N^{33}_{k'l'}
 \la 0|  a_{v}^{r'}   \, a^i_{k'}a^i_{l'} 
 \, a^j_{-k}a^j_{-l}   a_{-u}^r  |0\ra \ , \ea
where we used the fact that only when the modes $a_{\pm 2}$ are contracted among themselves we obtain a leading
contribution. 
Performing the contractions between the modes in the vacuum expectation value in the second line of the previous equation
 and taking into account the symmetry
of the coefficients $N^{33}_{kl}$ with respect to the exchange of the lower indices we find
\ba &&  {\rm Im}{\cal A}_{L L}  \sim
\frac{\pi}{2} \ap \b^2 \,  \si ( \z \e \de ) \si ( \bar \z \bar \e \de)  \sum_{u, k, l \atop
v, k', l'} \, \d_{u+k+l, n} N^{23}_{2u} N^{23}_{2v}N^{33}_{kl}N^{33}_{k'l'}
\si (48  klu \d_{u v}\d_{k k'} \d_{l l'} + 4 u v l \d_{u k} \d_{v k'} \d_{l l'}\de ) \nb \\
&=&\frac{\pi}{2}  \ap \b^2 \,  \si ( \z \e \de ) \si ( \bar \z \bar \e \de) \sum_{u, k, l} \, \d_{u+k+l, n} \, k l  u 
\si [48   \si(N^{23}_{2u}N^{33}_{kl}\de)^2 + 4 
N^{23}_{2u} N^{23}_{2k}N^{33}_{ul}N^{33}_{kl}  \de ]  \\
&=& \pi \ap \b^2 \,  \si ( \z \e \de ) \si ( \bar \z \bar \e \de) \, \sum_{u, k, l} \, \d_{u+k+l, n} \, k l  u 
\, \frac{\si(\s_1(u)\s_1(k)\s_1(l)\de)^2}{\si(2-\frac{u}{n}\de)(k+l)}
\si [  \frac{96 }{\si(2-\frac{u}{n}\de)(k+l)} 
+  \frac{8}{\si(2-\frac{k}{n}\de)(u+l)}
\de ] \nb \ . \ea
We now approximate the sums with integrals setting $u=n x_1$, $k=n x_2$, $l=n x_3$ and taking the large $n$ limit,
as discussed in Section \ref{heco}
\ba &&   {\rm Im}{\cal A}_{L L}  \sim \pi \ap \b^2 \,  n \,  \si ( \z \e \de ) \si ( \bar \z \bar \e \de)  \si [96  I_{L, 1}  + 8 I_{L, 2} \de ] \ , \ea
where 
\ba I_{L, 1} &=&  \int d\m_3 \,  x_1 x_2 x_3  \si(\frac{\s_1(x_1)\s_1(x_2)\s_1(x_3)}{\si(2-x_1\de)(x_2+x_3)}\de)^2 \sim 0.0082007 \ ,  \\
I_{L, 2} &=&  \int d\m_3 \,    \frac{x_1 x_2 x_3 \si(\s_1(x_1)\s_1(x_2)\s_1(x_3)\de)^2}{\si(2-x_1\de)\si(2-x_2\de)(x_1+x_3)(x_2+x_3)}
\sim 0.0066155 \ . \nb \ea
The second contribution is proportional to
\ba && \frac{1}{64} N^{23}_{2u}N^{23}_{2v}N^{33}_{kl}N^{33}_{gh}N^{33}_{k'l'}N^{33}_{g'h'} 
 \la 0| a_2^{s'} a_{v}^{r'}   \, a^i_{k'}a^i_{l'}   \, a^j_{g'}a^j_{h'} 
 \, a^w_{-g}a^w_{-h}   a_{-k}^q  a_{-l}^q a^r_{-u} a_{-2}^{s}|0\ra  \\
&=&  \frac{1}{32} N^{23}_{2u}N^{23}_{2v}N^{33}_{kl}N^{33}_{gh}N^{33}_{k'l'}N^{33}_{g'h'}  \, \d^{ss'} \si[ 
4 u g \d^{rr'} \d^{iw} \d_{uv} \d_{gk'} + 16 u g  \d^{wr'} \d^{ir} \d_{gv} \d_{uk'}\de]\, 
 \la 0| a^i_{l'}   \, a^j_{g'}a^j_{h'}  \, a^w_{-h}   a_{-k}^q  a_{-l}^q |0\ra  \nb \\
&=&  \frac{1}{32} N^{23}_{2u}N^{23}_{2v}N^{33}_{kl}N^{33}_{gh}N^{33}_{k'l'}N^{33}_{g'h'}  \, \d^{rr'} \d^{ss'} ughkl\si( 
96 \d_{uv} \d_{gk'} + 16   \d_{gv} \d_{uk'}\de)\, \si(
48 \d_{l'h} \d_{g'k} \d_{h'l}+ 4   \d_{g'h} \d_{h'l}\d_{kl'}\de) \ ,  \nb \ea
where we left the sum understood. Setting $u=n x_1$, $g=n x_2$, $h=n x_3$, $k=n x_4$, $l=n x_5$ and taking the large $n$ limit
we find
\ba n \, \d^{rr'} \d^{ss'} \, \si ( 576 \, I_{L, 3} + 96 \, I_{L, 4} + 48 \,   I_{L, 5} + 8 \,   I_{L, 6} \de ) \ ,
\label{B.8}
 \ea
where
\ba
\label{B.9}
 I_{L, 3} &=& \int d\m_5 \,  x_1 x_2 x_3 x_4 x_5 
\si(\frac{\s_1(x_1)\s_1(x_2)\s_1(x_3)\s_1(x_4)\s_1(x_5)}{\si(2-x_1\de)(x_2+x_3)(x_4+x_5)}\de)^2 \sim 0.0002179\ , \\
 I_{L, 4} &=&\int d\m_5 \, 
\frac{x_1 x_2 x_3 x_4 x_5 \si(\s_1(x_1)\s_1(x_2)\s_1(x_3)\s_1(x_4)\s_1(x_5)\de)^2}{\si(2-x_1\de)\si(2-x_2\de)(x_2+x_3)(x_4+x_5)^2(x_1+x_3)} 
\sim  0.0001709 \ , \nb \\
 I_{L, 5} &=& \int d\m_5 \, 
\frac{x_1 x_2 x_3 x_4 x_5 \si(\s_1(x_1)\s_1(x_2)\s_1(x_3)\s_1(x_4)\s_1(x_5)\de)^2}{\si(2-x_1\de)^2(x_2+x_3)(x_2+x_4)(x_4+x_5)(x_3+x_5)} 
\sim 0.0002087 \ , \nb \\
 I_{L, 6} &=&\int d\m_5 \, 
\frac{x_1 x_2 x_3 x_4 x_5 \si(\s_1(x_1)\s_1(x_2)\s_1(x_3)\s_1(x_4)\s_1(x_5)\de)^2}{\si(2-x_1\de)\si(2-x_2\de)(x_2+x_3)(x_4+x_5)(x_1+x_4)(x_3+x_5)} 
\sim  0.0001667\ . \nb 
\ea
Adding the second contribution to the first we obtain
\be {\rm Im}{\cal A}_{L L}  \sim \pi \ap \b^2 \,  n \, 
 (\zeta \epsilon) ( {\bar{\zeta}}  {\bar{\epsilon}})
\si  (0.841+0.153 \de ) =\pi \ap \b^2 \,  n \, 
 (\zeta \epsilon) ( {\bar{\zeta}}  {\bar{\epsilon}})  \, 0.994 \ . 
\label{B.10}
\ee
We now give the details of the derivation of the higher order terms in the 
series expansion of the imaginary part of the disk for the state $|H, \bar H\ra$. We start by
rewriting  Eq. (\ref{5.69})
\ba && {\rm Im}{\cal A}_{H H}  \sim \pi \ap \b^2 \, \si( \hat\e_{ij} \hat\z_{ij} \de ) \frac{1}{8}\la 0|\si [ n^2 \,
\hat{\bar{\z}}_{ll}  + 2  N^{23}_{1u'} N^{23}_{1v'}  (D\hat{\bar{\z}}D)_{r's'}
 a_{u'}^{r'}  a_{v'}^{s'}    \de ]  e^{Z_o^\dagger} \nb \\
&& {\cal P}_n e^{Z_o}
\si [ n^2 \,
\hat{\bar{\e}}_{kk} + 2  N^{23}_{1u} N^{23}_{1v}   (D\hat{\bar{\e}}D)_{rs}
a_{-u}^r  a_{-v}^s    \de ]  |0\ra \ . \label{ppB} \ea 
In addition to the term already evaluated in Eq. (\ref{5.70}) of Section \ref{heco},
we have the following terms with two summation variables
\ba \pi \ap \b^2 \, \si(\hat\e_{ij} \hat\z_{ij} \de ) \frac{1}{8} \sum_{k, l} \d_{k+l,n}\si [ 4 kl N^{23}_{1k}N^{23}_{1l} N^{33}_{kl}
+12 n^2 kl \si( N^{33}_{kl} \de )^2\de ] \, n^2 \, \hat{\bar{\e}}_{rr}\hat{\bar{\z}}_{ss} \ , \ea
that in the continuum limit become
\ba 
\label{B.13}
\pi \ap \b^2 \, n \, \, \si( \hat\e_{ij} \hat\z_{ij} \de ) \sum_{k, l} \d_{k+l,n}\si [ - \frac{1}{2} I_{H, 2}
+ \frac{3}{2} I_{H, 3} \de ] \, \hat{\bar{\e}}_{rr}\hat{\bar{\z}}_{ss}\ , \ea
where
\ba I_{H, 2} &=& \int d\m_2 \, x_1 x_2 \frac{\si(\s_1(x_1)\s_1(x_2)\de)^2}{(1-x_1)(1-x_2)(x_1+x_2)} \sim 0.178692 \ , \nb \\ 
I_{H, 3} &=& \int d\m_2 \, x_1 x_2 \frac{\si(\s_1(x_1)\s_1(x_2)\de)^2}{(x_1+x_2)^2} \sim 0.039863 \ . \ea
In Eq. (\ref{B.13}) we have obtained two terms that do not have the correct contraction 
for the polarizations as discussed in Eq. (\ref{B.73}). Therefore, let us go back to Eq. (\ref{ppB})
and include also higher order  terms that come from the expansion of the two  exponentials
containing $Z_{o}$ and $Z_0^\dagger$. 
There are four terms with four summation variables that contribute to the imaginary part. 
The first one is  
\ba && \pi \ap \b^2 \, \si( {\hat{\e}}_{ij} {\hat{\z}}_{ij} \de )  (D\hat{\bar{\z}}D)_{r's'}  (D\hat{\bar{\e}}D)_{rs} \,  \frac{1}{8} N^{23}_{1u}N^{23}_{1v}N^{33}_{kl}
N^{23}_{1u'}N^{23}_{1v'}N^{33}_{k'l'} \la 0|a^{r'}_{u'}a^{s'}_{v'}a^{i}_{k'}a^{i}_{l'}a^{j}_{-k}a^{j}_{-l}a^{r}_{-u}a^{s}_{-v} |0\ra \nb \\
&=& \pi \ap \b^2 \, n \, \si( \hat\e_{ij} \hat\z_{ij} \de )  \si[ \si(12 I_{H, 4} + 2 I_{H, 5}\de) \si( \hat{\bar \e}_{kl} \hat{\bar \z}_{kl}\de)+\frac{1}{2}
I_{H, 6} \, \hat{\bar{\e}}_{kk}\hat{\bar{\z}}_{ll} \de ]\ , \ea 
where the three integrals correspond to three inequivalent contractions of the modes in the vacuum expectation value
and are given by 
\ba I_{H, 4} &=& \int d\m_4 \, x_1 x_2 x_3 x_4  \frac{\si(\s_1(x_1)\s_1(x_2)\s_1(x_3)\s_1(x_4)\de)^2}{(1-x_1)^2(1-x_2)^2(x_3+x_4)^2} \sim
 0.012072 \ ,  \\
I_{H, 5} &=& \int d\m_4 \, x_1 x_2 x_3 x_4 
\frac{\si(\s_1(x_1)\s_1(x_2)\s_1(x_3)\s_1(x_4)\de)^2}{(1-x_1)^2(1-x_2)(1-x_3)(x_3+x_4)(x_2+x_4)} \sim  0.007971 \ , \nb \\
I_{H, 6} &=& \int d\m_4 \, x_1 x_2 x_3 x_4 
\frac{\si(\s_1(x_1)\s_1(x_2)\s_1(x_3)\s_1(x_4)\de)^2}{(1-x_1)(1-x_2)(1-x_3)(1-x_4)(x_1+x_2)(x_3+x_4)} \sim 0.005733 \nb \ . \ea
The second is
\ba && \pi \ap \b^2 \, \si( \hat\e_{ij} \hat\z_{ij} \de )  \hat{\bar{\z}}_{ss}\, (D\hat{\bar{\e}}D)_{r's'} \,  \frac{n^2}{64} N^{23}_{1u}N^{23}_{1v}N^{33}_{kl}
N^{33}_{k'l'}N^{33}_{gh}\, \la 0|a^{i}_{k'}a^{i}_{l'}a^{j}_{g}a^{j}_{h}a^{w}_{-k}a^{w}_{-l}a^{r'}_{-u}a^{s'}_{-v} |0\ra \nb \\
&=& - \pi \ap \b^2\, n  \, \si( \hat\e_{ij} \hat\z_{ij} \de )  \, \hat{\bar{\e}}_{rr}\hat{\bar{\z}}_{ss}
  \si[3 I_{H, 7} +  \frac{1}{4}I_{H, 8}\de ]\ , \ea 
where the two integrals are
\ba I_{H, 7} &=& \int d\m_4 \, x_1 x_2 x_3 x_4  \frac{\si(\s_1(x_1)\s_1(x_2)\s_1(x_3)\s_1(x_4)\de)^2}{(1-x_1)(1-x_2)(x_1+x_2)(x_3+x_4)^2} \sim
0.008181 \ ,  \\
I_{H, 8} &=& \int d\m_4 \, x_1 x_2 x_3 x_4 
\frac{\si(\s_1(x_1)\s_1(x_2)\s_1(x_3)\s_1(x_4)\de)^2}{(1-x_1)(1-x_2)(x_3+x_4)(x_1+x_3)(x_2+x_4)} \sim    0.007700 \ \nb \ . \ea
The third gives a contribution identical to the second  and the fourth is 
\ba && \pi \ap \b^2 \, \si( \hat\e_{ij} \hat\z_{ij} \de )  \, \hat{\bar{\e}}_{rr}\hat{\bar{\z}}_{ss} \,  \frac{n^4}{512} N^{33}_{kl}N^{33}_{gh}
N^{33}_{k'l'}N^{33}_{g'h'}\, \la 0|a^{i}_{k'}a^{i}_{l'}a^{j}_{g'}a^{j}_{h'}a^{w}_{-k}a^{w}_{-l}a^{t}_{-g}a^{t}_{-h} |0\ra \nb \\
&=&  \pi \ap \b^2 \, n \, \si( \hat\e_{ij} \hat\z_{ij} \de ) \, \hat{\bar{\e}}_{rr}\hat{\bar{\z}}_{ss}
  \si[9 I_{H, 9} +  \frac{3}{4}I_{H, 10}\de ]\ , \ea 
where the two integrals are
\ba I_{H, 9} &=& \int d\m_4 \, x_1 x_2 x_3 x_4  \frac{\si(\s_1(x_1)\s_1(x_2)\s_1(x_3)\s_1(x_4)\de)^2}{(x_1+x_2)^2(x_3+x_4)^2} \sim
0.007789 \ ,  \\
I_{H, 10} &=& \int d\m_4 \, x_1 x_2 x_3 x_4 
\frac{\si(\s_1(x_1)\s_1(x_2)\s_1(x_3)\s_1(x_4)\de)^2}{(x_1+x_2)(x_3+x_4)(x_1+x_3)(x_2+x_4)} \sim    0.007538 \ \nb \ . \ea
Collecting all the terms with two and four summation variables we then find
\ba && {\rm Im}{\cal A}_{H H}  \sim \pi \ap \b^2 \, n \, \si( \hat\e_{ij} \hat\z_{ij} \de ) \si [ \si (I_{H,1} + 12  I_{H, 4}  + 2  I_{H, 5}  \de )
\si( \hat{\bar \e}_{kl} \hat{\bar \z}_{kl} \de ) \de . \nb \\
&+& \si .  \si(  -  \frac{1}{2}  I_{H, 2} + \frac{3}{2}   I_{H, 3}   +   \frac{1}{2} I_{H, 6}  - 6  I_{H, 7}  -
 \frac{1}{2}  I_{H, 8}  + 9  I_{H, 9}  + \frac{3}{4}  I_{H, 10} \de)  \, \hat{\bar{\e}}_{kk}\hat{\bar{\z}}_{ll} \de] \nb \\
&\sim& \pi \ap \b^2 \, n \, \si( \hat\e_{ij} \hat\z_{ij} \de ) \si [ 0.993 \si( \hat{\bar \e}_{kl} \hat{\bar \z}_{kl} \de ) 
- 0.004 \, \hat{\bar{\e}}_{kk}\hat{\bar{\z}}_{ll} \de] \ . \ea
The terms contributing to the first kind of contraction are all positive and seem to add to one. The sign of the terms contributing to the
second kind of contraction alternates with the number of factors of $N^{33}$ and seem to add to zero.  In fact the result
of the sum is one order of  magnitude smaller than the individual terms.

\end{appendix}

\end{document}